\RequirePackage{silence}
\documentclass[fleqn,usenatbib,useAMS]{mnras} %Class file altered on lines 116, 119, 122 and 123.
\usepackage{graphicx}
\usepackage{amsmath}

\usepackage{amsfonts}
\usepackage{float}
\usepackage{bm}
\usepackage[perpage,symbol*]{footmisc}%At end of square bracket: ,symbol* \usepackage[T1]{fontenc}
\usepackage{ae,aecompl}
\usepackage{array}
\usepackage{soul}
\usepackage{mathtools}
\usepackage{multirow}

\usepackage{color}
\usepackage{xcolor}
\definecolor{msp}{rgb}{0.0,0.6,0.6}

\usepackage{ulem} % to strike out text with \sout{text}

\title[NGC 3109 is not a backsplash galaxy in $\Lambda$CDM]{On the absence of backsplash analogues to NGC 3109 in the $\Lambda$CDM framework} 

\author[I. Banik et al.]{\parbox[t]{\textwidth} {Indranil Banik$^{1}$\thanks{Email:
\href{mailto:ibanik@astro.uni-bonn.de}{ibanik@astro.uni-bonn.de} (Indranil Banik)\newline $~~~~~~~~\,$ \href{mailto:s6mohasl@astro.uni-bonn.de}{mhaslbauer@astro.uni-bonn.de} (Moritz Haslbauer)}, Moritz Haslbauer$^{1,2}$, Marcel S. Pawlowski$^3$, Benoit Famaey$^4$ \& Pavel Kroupa$^{1,5}$} \vspace{10pt} \\
$^{1}$Helmholtz-Institut f\"ur Strahlen und Kernphysik (HISKP), University of Bonn, Nussallee 14$-$16, D-53115 Bonn, Germany \\
$^{2}$Max-Planck-Institut f\"ur Radioastronomie, Auf dem H\"ugel 69, D-53121 Bonn, 
Germany \\
$^{3}$Leibniz-Institut f\"ur Astrophysik Potsdam (AIP), An der Sternwarte 16, D-14482 Potsdam, Germany \\
$^{4}$Universit\'{e} de Strasbourg, CNRS UMR 7550, Observatoire astronomique de Strasbourg, 11 rue de l'Universit\'{e}, 67000 Strasbourg, France \\
$^{5}$Astronomical Institute, Faculty of Mathematics and Physics, Charles University, V Hole\v{s}ovi\v{c}k\'ach 2, CZ-180 00 Praha 8, Czech Republic}

\pubyear{2021}
\begin{document}
\label{firstpage}
\pagerange{\pageref{firstpage}--\pageref{lastpage}}

\maketitle

\begin{abstract}

The dwarf galaxy NGC 3109 is receding 105 km/s faster than expected in a $\Lambda$CDM timing argument analysis of the Local Group and external galaxy groups within 8 Mpc (Banik \& Zhao 2018). If this few-body model accurately represents long-range interactions in $\Lambda$CDM, this high velocity suggests that NGC 3109 is a backsplash galaxy that was once within the virial radius of the Milky Way and was slingshot out of it. Here, we use the Illustris TNG300 cosmological hydrodynamical simulation and its merger tree to identify backsplash galaxies. We find that backsplashers as massive ($\geq 4.0 \times 10^{10} M_\odot$) and distant ($\geq 1.2$ Mpc) as NGC 3109 are extremely rare, with none having also gained energy during the interaction with their previous host. This is likely due to dynamical friction. Since we identified 13225 host galaxies similar to the Milky Way or M31, we conclude that postulating NGC 3109 is a backsplash galaxy causes $>3.96\sigma$ tension with the expected distribution of backsplashers in $\Lambda$CDM. We show that the dark matter only version of TNG300 yields much the same result, demonstrating its robustness to how the baryonic physics is modelled. If instead NGC 3109 is not a backsplasher, consistency with $\Lambda$CDM would require the 3D timing argument analysis to be off by 105 km/s for this rather isolated dwarf, which we argue is unlikely. We discuss a possible alternative scenario for NGC 3109 and the Local Group satellite planes in the context of MOND, where the Milky Way and M31 had a past close flyby $7-10$ Gyr ago.

\end{abstract}


\section{Introduction}
\label{Introduction}

The Universe was expanding rather homogeneously at early times \citep{Planck_2014}, yet the present velocities of galaxies in the Local Group (LG) deviate significantly from a pure Hubble flow. This is due to the gravity they exert on each other. However, the large distance between the Milky Way (MW) and Andromeda (M31) galaxies implies only a rather weak gravitational attraction if we consider the Newtonian gravity of their baryons alone. This is insufficient to turn around their initial expansion and cause them to approach each other at the observed rate of ${\approx 110}$~km/s \citep{Van_der_Marel_2012}. Clearly, there must be an extra source of gravity between the MW and M31. This `timing argument' was one of the oldest arguments for missing gravity on galactic scales \citep{Kahn_Woltjer_1959}.

The now standard Lambda-Cold Dark Matter ($\Lambda$CDM) cosmological paradigm \citep{Efstathiou_1990, Ostriker_Steinhardt_1995} proposes that galaxies like the MW and M31 formed and evolved within haloes \citep{WhiteRees78} of non-baryonic cold dark matter (CDM) particles, accounting for the missing gravity. In this framework, the CDM haloes of the MW and M31 must be massive enough to turn around their initial expansion to the observed extent within the available 13.8~Gyr. This constrains their total mass \citep[e.g.][]{Carlesi_2017}. However, this is not a strong test of the $\Lambda$CDM model because there are as many data points as model parameters $-$ the total mass and initial separation are varied to match the present separation and radial velocity (RV). This degeneracy can be broken by including data on more distant galaxies in the LG. Such an analysis was attempted by \citet{Sandage_1986}, who found it difficult to simultaneously explain all the data then available. Using the catalogue of LG dwarfs in \citet{McConnachie_2012}, a similar study was attempted by \citet{Jorge_2014}, who found that the observations require an additional source of uncertainty with magnitude ${35 \pm 5}$~km/s.

Following on from these spherically symmetric dynamical models, \citet{Banik_Zhao_2016} constructed an axisymmetric model of the LG consistent with the almost radial MW-M31 orbit \citep{Van_der_Marel_2012} and the close alignment of Centaurus A with this line \citep{Ma_1998}. The nearly radial nature of the MW-M31 orbit was later confirmed by \citet{Van_der_Marel_2019} and \citet{Salomon_2021}, which will be important to our discussion in Section \ref{MOND_scenario}. Treating LG dwarfs as test particles in the gravitational field of these three massive moving objects, \citet{Banik_Zhao_2016} investigated a wide range of model parameters using a full grid search. None of the models produced a good fit, even when reasonable allowance was made for inaccuracies in the model as a representation of $\Lambda$CDM based on the scatter about the Hubble flow in detailed $N$-body simulations \citep{Aragon_Calvo_2011}. This is because several LG dwarfs have receding RVs much higher than expected in the best-fitting model, though the opposite is rarely the case \citep[figure 9 of][]{Banik_Zhao_2016}.

\citet{Banik_Zhao_2017} used the algorithm described in \citet{Shaya_2011} to test whether this remains the case when using a 3D model of the LG. The typical mismatch between observed and predicted RVs in the best-fitting model is even higher than in the 2D case, with a clear tendency persisting for faster outward motion than expected \citep[figures 7 and 9 of][]{Banik_Zhao_2017}. These results are comparable to those obtained by \citet{Peebles_2017} using a similar algorithm. \citet{Banik_2018_anisotropy} borrowed this algorithm from Peebles and made some significant improvements in order to maximize the chance of finding trajectories consistent with the timing argument (see their section 4.1). Nevertheless, the results remained almost unchanged, with the only major difference being that Tucana became consistent with the model. An important clue is that nearly all the high-velocity galaxies (HVGs) are part of the NGC~3109 association, which was previously identified as having properties that are difficult to understand in $\Lambda$CDM \citep{Pawlowski_McGaugh_2014}. The heliocentric RV of NGC~3109 is 403~km/s, which translates to 170 km/s in the Galactocentric frame, slightly below the expected value for a pure Hubble flow (without gravity) centred on the LG barycentre. However, taking into account the effect of Newtonian gravity, this is still 105~km/s too high in the best-fitting model \citep{Banik_2018_anisotropy}.

Such a high RV reduces the LG timing argument mass inferred from the kinematics of non-satellite dwarf galaxies outside the MW and M31 virial volumes. This might well explain the unusually low LG mass of $\left( 1.6 \pm 0.2 \right) \times 10^{12} M_\odot$ found in this manner by \citet{Kashibadze_2018}, with their table~4 indicating that their analysis included the NGC~3109 association. \citet{Zhai_2020} obtained a much higher timing argument mass of $4.4^{+2.4}_{-1.5} \times 10^{12} M_\odot$ by searching cosmological simulations for analogues to the LG based on properties of the MW and M31 alone, especially with regards to their relative separation and velocity. This mass is in line with earlier results and simple analytic estimates neglecting information on LG galaxies other than the MW and M31 \citep{Li_White_2008}. The mass of M31 alone has been estimated at $1.9_{-0.4}^{+0.5} \times 10^{12} M_\odot$ based on its giant southern tidal stream \citep{Fardal_2013}. The total LG mass is certainly higher as it also includes the MW and material outside the major LG galaxies. Thus, several timing argument analyses of the whole LG found it difficult to explain the high RVs of some dwarf galaxies, with the tension phrased in some works as an anomalously low LG timing argument mass.

The timing argument calculations in \citet{Peebles_2017} and \citet{Banik_2018_anisotropy} are however not perfect representations of $\Lambda$CDM. They should handle long-range interactions between galaxies rather well, but can potentially miss important details due to the lack of dynamical friction between DM haloes. They consequently lack simulated mergers, during which galaxies can temporarily have a high relative velocity that could slingshot a nearby dwarf outwards at high speed in a three-body interaction. This leads to the existence of so-called `backsplash galaxies' (backsplashers), defined as objects on rather extreme orbits that were once within the virial radius of their host but were subsequently carried outside of it. This backsplash process was studied in detail by \citet{Sales_2007}, who found it very difficult to get backsplashers at the $1.30 \pm 0.02$~Mpc distance of NGC~3109 \citep{Soszynski_2006}. This is also evident in figure~3 of \citet{Teyssier_2012}, which additionally showed that backsplashers are very rarely more massive than $10^{10} M_\odot$ regardless of their present position. Though NGC 3109 is more massive (Section \ref{NGC3109_mass}), they argued that it is most likely a backsplasher given its position and RV. In a $\Lambda$CDM context, it might be very difficult to obtain such a massive and distant backsplasher due to the expected dynamical friction during any encounter with the MW or M31 \citep[e.g. section 4.2 of][]{Kroupa_2015}. This would entail ejecting a galaxy as massive as NGC 3109 out of the inner regions of the MW halo against the inevitable dynamical friction. However, a more distant interaction with less dynamical friction would be rather weak, thus having little effect on the trajectory of NGC 3109 beyond that included in a few-body model.

In this contribution, we revisit the $\Lambda$CDM-predicted distribution of dwarfs around analogues to the MW or M31 using the much larger volume of the Illustris TNG300 cosmological hydrodynamical simulation \citep{Pillepich_2018}. We also use the corresponding dark matter-only simulation to check how our results are affected by modelling of the baryonic physics. Our main goal is to find simulated backsplashers with NGC~3109-like properties today, but whose trajectories are likely to be seriously mis-modelled in the few-body analyses of \citet{Peebles_2017} and \citet{Banik_2018_anisotropy}. If no such trajectories exist, this would improve our confidence in how well those models represent the underlying $\Lambda$CDM paradigm, thereby confirming the challenge posed by NGC~3109.

In Section~\ref{Simulation}, we describe the essential characteristics of the simulation for the present work. We then review the observed properties of NGC~3109 and our selection criteria in Section~\ref{Observed_properties_NGC3109}. In Section~\ref{NGC3109analogs}, we search for analogues of NGC~3109 in the simulation without requiring it to be a backsplasher, and review the timing argument analysis for the observed LG and its expected reliability. We then search for analogue backsplashers in Section~\ref{NGC3109abacksplash}, and present the results in Section~\ref{Results}. We discuss our results and an alternative scenario in Section~\ref{Discussion}, and conclude in Section~\ref{Conclusions}.

\section{Cosmological simulation}
\label{Simulation}

To explore whether the few-body models of \citet{Peebles_2017} and \citet{Banik_2018_anisotropy} might miss trajectories that explain the anomalous kinematics of NGC~3109 within a $\Lambda$CDM framework, we use the Illustris TNG300-1 hydrodynamical cosmological simulation \citep{Pillepich_2018, Nelson_2019}. This investigates $\Lambda$CDM in a cubic region with side length $205 \, h^{-1}$ co-moving Mpc (cMpc). The Hubble constant $H_0$ in units of 100~km/s/Mpc is $h = 0.6774$, so the simulation box has a side length of 302.6~cMpc (hence the name TNG300). The suffix `-1' indicates that we use the highest available resolution setting for this box size within the Illustris suite. These simulations assume a standard flat cosmology in which the present fraction of the cosmic critical density in matter is 0.3089 \citep{Planck_2016}. The low $H_0$ in this cosmology is also required by the early Universe observations of \citet{Aiola_2020}.

We use TNG300 because the larger simulation box compared to TNG100 or TNG50 allows for better statistics. All these simulations can adequately resolve objects much less massive than NGC~3109, as will become apparent in Section \ref{Results}. However, we expect that we must search through many host galaxies analogous to the MW or M31 to find any backsplashers with properties similar to NGC~3109, or to set a stringent upper limit on their occurrence rate. In what follows, we will refer to host galaxies simply as `MW analogues' even though the allowed range of properties are extended to allow M31-like galaxies, leaving open the possibility that NGC~3109 is backsplash from M31 (Section \ref{Isolation_conditions}). However, we will see that this is much less plausible than backsplash from the MW.

The Illustris catalogues contain 100 snapshots going back from the present epoch to when the cosmic scale factor was $a = 0.0475$. The catalogues distinguish between groups and subhaloes. We use the redshift $z = 0$ group catalogue to identify isolated or LG-like host galaxies, though with some additional checks based on the $z = 0$ subhalo catalogue (Section \ref{Isolation_conditions}). The position of the MW analogue at any epoch is found using the subhalo catalogue for that epoch, while its virial radius is found using the group catalogue as this is the only one to list virial radii. We use the subhalo catalogues to obtain properties of candidate backsplashers at various epochs. The Illustris \textsc{sublink} merger tree \citep{Gomez_2015} allows us to trace back MW analogues and to trace forward subhaloes within their virial volume, and finally to trace back any candidate backsplasher to better understand its trajectory.

We start by compiling in Section~\ref{Observed_properties_NGC3109} the observed properties of NGC~3109, and our implemented criteria when selecting analogues in the Illustris TNG300 simulation. In Section~\ref{NGC3109analogs}, we search for analogues of NGC~3109 in this simulation without requiring it to be a backsplasher, followed by a review of the timing argument analysis in the detailed context of the LG. The rest of this paper concerns the backsplash analysis of the Illustris TNG300 simulation.

\section{Observed properties of NGC 3109}
\label{Observed_properties_NGC3109}

To select simulated galaxies analogous to NGC~3109, we first review its observed properties. Due to its small distance and fairly high mass, the uncertainties are rather small.

\subsection{Distance}
\label{NGC3109_distance}

The Galactocentric distance of NGC~3109 is one of the most important observational inputs to our analysis. It was measured to be $1.30 \pm 0.02$~Mpc \citep{Soszynski_2006}. Similar results were obtained by \citet{Dalcanton_2009} and several other studies. To be conservative, we adopt a distance at the $5\sigma$ lower limit of the observationally allowed range. Thus, we require that analogues to NGC~3109 be $\geq 1.2$~Mpc from their host.

It is possible that NGC~3109 is a backsplasher from M31, whose merger history appears to have been more active than that of the MW \citep[e.g.][]{Hammer_2010, Souza_2018}. According to table 2 of \citet{McConnachie_2012}, the separation between NGC~3109 and M31 is currently 1.99~Mpc. The larger distance arises because the whole NGC~3109 association is in the opposite hemisphere on our sky compared to M31 \citep[e.g. see figure 16 of][]{Banik_2018_anisotropy}. We will see later that this makes a backsplash event in M31 a much less plausible scenario than backsplash from the MW. Thus, we focus almost exclusively on the hypothesis that NGC~3109 was once within the virial radius of the MW in a $\Lambda$CDM context. For Illustris host galaxies in a paired LG-like configuration, we require that the backsplasher lies $\geq 1.2$~Mpc from both hosts.

\subsection{Mass}
\label{NGC3109_mass}

\begin{figure}
	\centering
	\includegraphics[width = 8.5cm] {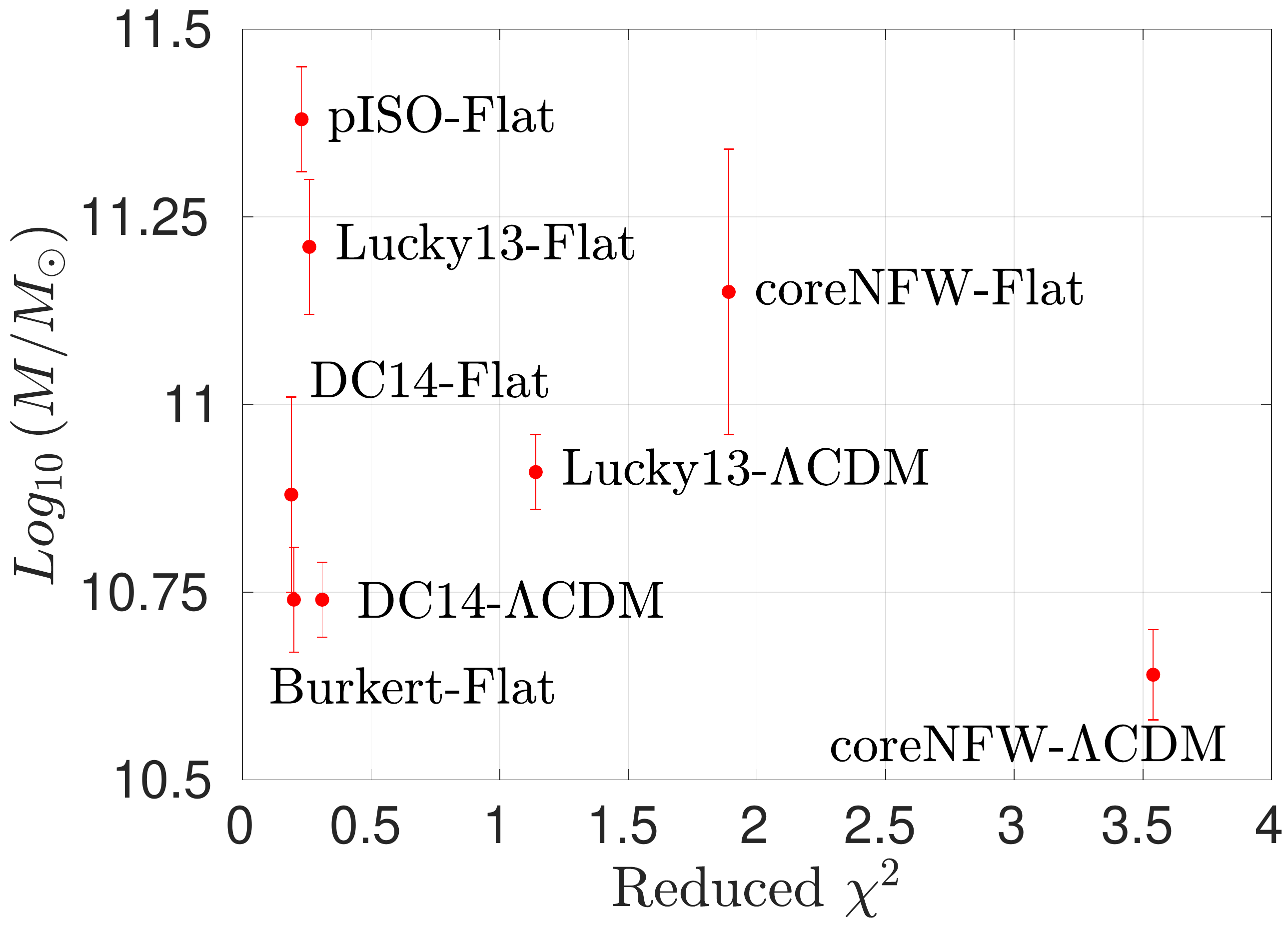}
	\caption{The estimated virial mass of NGC~3109 according to Newtonian rotation curve fits with different halo profiles (text labels), shown against the reduced $\chi^2$ of the model if this is $<9$ \citep{Li_2020}. Based on these results, we conservatively assume that NGC~3109 is more massive than $10^{10.6} M_\odot = 4.0 \times 10^{10} M_\odot$.}
	\label{Li_results}
\end{figure}

In addition to the large distance of NGC~3109, we also need to explain its large mass for a backsplasher in the $\Lambda$CDM framework. Its virial mass can be estimated using rotation curve fits that add various halo profiles to the observed baryonic components. Several such fits were conducted by \citet{Li_2020} based on the Spitzer Photometry and Accurate Rotation Curves dataset \citep[SPARC;][]{SPARC}. The results are summarized in Figure \ref{Li_results}, which shows only those halo profiles with a reduced $\chi^2$ below 9. Based on these results, we conservatively assume that the mass of NGC~3109 is at least $10^{10.6} M_\odot = 4.0 \times 10^{10} M_\odot$, since no acceptable fits were obtained with a lower mass. Several studies give larger values, with \citet{Valenzuela_2007} estimating a virial mass of $8.1 \times 10^{10} M_\odot$ (see their table 4). Moreover, the total mass (as recorded in the Illustris catalogue) is expected to exceed the virial mass \citep{Jorge_2017}. For our purposes, using a lower mass is more conservative as we expect less massive subhaloes to be flung out further via the backsplash process, making it easier to match the large distance of NGC~3109.

These mass estimates do not consider the rest of the NGC~3109 association, which consists of galaxies at a similar position with a similar anomalously high RV \citep{Pawlowski_McGaugh_2014}. As discussed in Section \ref{Discussion} and in \citet{Banik_2018_anisotropy}, it is very unlikely that multiple dwarf galaxies were flung out in the same direction at a similar time despite initially being independent of each other. A more plausible explanation is that they formed a gravitationally bound association, though this is likely not bound any more \citep{Kourkchi_2017}. If so, the mass of NGC~3109 must be much higher, with \citet{Bellazzini_2013} estimating a mass of $3.2 \times 10^{11} M_\odot$. The relatively low stellar fraction (see below) could be due to tidal or ram pressure effects during a past interaction with the MW. Note that a past interaction could have resulted in loss of dark matter, so this estimate should be compared with the pre-interaction mass of each backsplasher. A conservative approach would be to compare with the maximum mass of each backsplasher at any snapshot in the Illustris simulation, which we discuss in Section \ref{Discussion}.

Since Illustris is a hydrodynamical simulation, we may instead compare the baryonic mass of each backsplasher to that of NGC~3109. This was estimated at $2.1 \times 10^9 M_\odot$ by applying rotation curve fitting techniques to high-resolution $N$-body models \citep[table 4 of][]{Valenzuela_2007}. Their estimated neutral hydrogen mass of $\left( 6 - 8 \right) \times 10^8 M_\odot$ is similar to the $\left( 4.6 \pm 0.5 \right) \times 10^8 M_\odot$ reported in table 8 of \citet{Carignan_2013}. Using stellar population synthesis modelling, \citet{Valenzuela_2007} estimated that $\approx 5 \times 10^8 M_\odot$ is contributed by stars, with the rest coming from gas.

The more recent SPARC database gives a 3.6~$\mu$m luminosity for NGC~3109 of $\left( 1.94 \pm 0.02 \right) \times 10^8 L_\odot$, which suggests a stellar mass of only $1.0 \times 10^8 M_\odot$ for a mass to light ratio of 0.5 \citep{Schombert_2014}. Combining this with $1.33 \times$ their estimated neutral hydrogen mass of $4.77 \times 10^8 M_\odot$ to account for primordial helium, we get a minimum possible baryonic mass of $7.3 \times 10^8 M_\odot$.

\begin{table}
	\centering
	\begin{tabular}{cc}
		\hline
		Component of NGC 3109 & Mass ($M_\odot$) \\
		\hline
		Stars & $1.0 \times 10^8$ \\
		Baryons (in disc) & $7.3 \times 10^8$ \\
		Virial (minimum) & $4.0 \times 10^{10}$ \\
		Virial (maximum) & $3.2 \times 10^{11}$ \\
		\hline
	\end{tabular}
	\caption{Parameters of NGC~3109 used in this study. The different virial mass estimates refer to whether we consider the kinematics of NGC~3109 alone, or require it and its associated galaxies to be gravitationally bound \citep{Bellazzini_2013}. We assume a Galactocentric distance of 1.2~Mpc to be conservative (Section \ref{NGC3109_distance}).}
	\label{NGC_3109_parameters}
\end{table}

Table \ref{NGC_3109_parameters} summarizes our adopted mass estimates for NGC~3109, where we have erred on the low side to be conservative. We focus on comparing the virial mass estimate with the total subhalo mass in Illustris backsplashers, as this should be least affected by uncertainties regarding subgrid baryonic feedback processes. When using the total mass, we still require that analogues to NGC~3109 have a non-zero stellar mass and thus a non-zero baryonic mass (for safety, we require both). We show later that our results remain much the same if we select backsplash analogues to NGC~3109 based on its stellar or baryonic mass.

\subsection{Isolation conditions and host properties}
\label{Isolation_conditions}

We consider two kinds of host galaxy $-$ LG-like and isolated. In both cases, we identify appropriate hosts by considering the $z = 0$ group catalogue based on the Friends of Friends (FoF) approach. We require each FoF group to have at least 1 subhalo. If it has $\geq 2$ subhaloes, we require the second most massive subhalo to comprise ${<20\%}$ of the group virial mass \citep[section 4.1 of][]{Pawlowski_2020}.

LG-like hosts consist of two FoF groups with a separation of $\left( 0.75 - 1.5 \right)$~Mpc and total virial mass of $\left( 2 - 5 \right) \times 10^{12} M_\odot$. This mass range covers the LG mass estimated in various ways, e.g. it is similar to the range reported by \citet{Gonzalez_2014} and \citet{Zhai_2020} based on LG analogues in cosmological simulations with a similar separation and relative velocity to the MW-M31 system. The 1D timing argument analyses of \citet{Jorge_2014} and \citet{Jorge_2016} give values near the lower end of this range. To have a reasonable mass ratio between the galaxies, we require that
\begin{eqnarray}
	\frac{M_{max}}{M_{min}} ~<~ 3 \, ,
	\label{Max_mass_ratio}
\end{eqnarray}
where $M_{max} \left( M_{min} \right)$ is the virial mass of the heavier (lighter) member of the candidate pair.

The lower limit on their separation is based on the 783~kpc distance to M31 \citep{McConnachie_2012}. Pairs with such a large separation are unlikely to have turned around and undergone an interaction within a Hubble time. Thus, requiring a present separation $>750$~kpc implicitly imposes the condition that the MW and M31 did not undergo a past close interaction, which is correct in a $\Lambda$CDM context given their nearly radial orbit \citep{Van_der_Marel_2012, Van_der_Marel_2019, Salomon_2021} and the consequent very strong dynamical friction in any close encounter \citep{Privon_2013}. Even without this consideration, a past flyby in Newtonian gravity would entail a very high timing argument mass \citep{Benisty_2019}. Including any LG analogues in Illustris which had such an interaction could seriously compromise our analysis as there could be backsplashers from the interaction, which as argued above would not be a viable scenario in $\Lambda$CDM. The upper limit on the separation prevents interference from neighbouring groups beyond 3 Mpc (see below).

We ensure a sufficient level of isolation by requiring there to be no other group within 3~Mpc that is more massive than $M_{min}/3$. We also remove pairs where there is another group more massive than $5 \left( M_{max} + M_{min} \right)$ within 5~Mpc, with the latter condition based on table 3 of \citet{Banik_Zhao_2017}. This avoids massive nearby groups interfering with the dynamics of the LG, e.g. by pulling a backsplasher out to a much greater distance than it would otherwise reach.

For consistency with the above criteria, we require isolated hosts to have a virial mass $M$ in the range $\left( 0.5 - 3.75 \right) \times 10^{12} M_\odot$. Their isolated nature is assured by requiring there to be no other group more massive than $M/3$ within 3~Mpc or more massive than $5M$ within 5~Mpc.

These selection criteria yield 13225 host galaxies, of which 640 are found in 320 LG-like paired configurations.

\section{NGC 3109 analogues in TNG300 and the timing argument}
\label{NGC3109analogs}

\subsection{Frequency of NGC~3109 analogues ignoring the detailed environment of the LG}
\label{Hubble_diagram_analysis}

We now start our analysis by determining the RVs of nearby galaxies with respect to our selected hosts after imposing all the conditions compiled in the previous section (in Section \ref{Backsplash_condition} we will add to this the backsplash condition). This allows us to focus on the observed properties of galaxies without assumptions about their dynamical history or status as a backsplasher, thereby ignoring for now the detailed observed environment of the LG.

Since the RV will depend on distance, we restrict to a narrow distance range of $\left( 1.1 - 1.5 \right)$~Mpc, which is wide enough to allow good statistics but narrow enough that there is little trend in RV with distance. The selected range is centred on the observed 1.30~Mpc distance to NGC~3109 \citep{Soszynski_2006}, allowing also a $\pm 10 \sigma$ uncertainty. For LG-like hosts, we only consider the above-mentioned distance range relative to the less massive galaxy, and require a distance $>1.5$~Mpc from the more massive galaxy. This resembles the LG somewhat more closely, though the statistics are dominated by isolated hosts.

%Backsplashers with valid GRV = 9121, of which 105 have GRV > 170 km/s, fraction = 0.011512.
%nanstd(v_r_rel_values) = 58.2923 km/s.
%
%Backsplashers with valid GRV = 9130, of which 171 have GRV > 105 km/s, fraction = 0.018729.
%nanstd(v_r_rel_values) = 46.7213 km/s.

\begin{figure}
	\centering
	\includegraphics[width = 8.5cm] {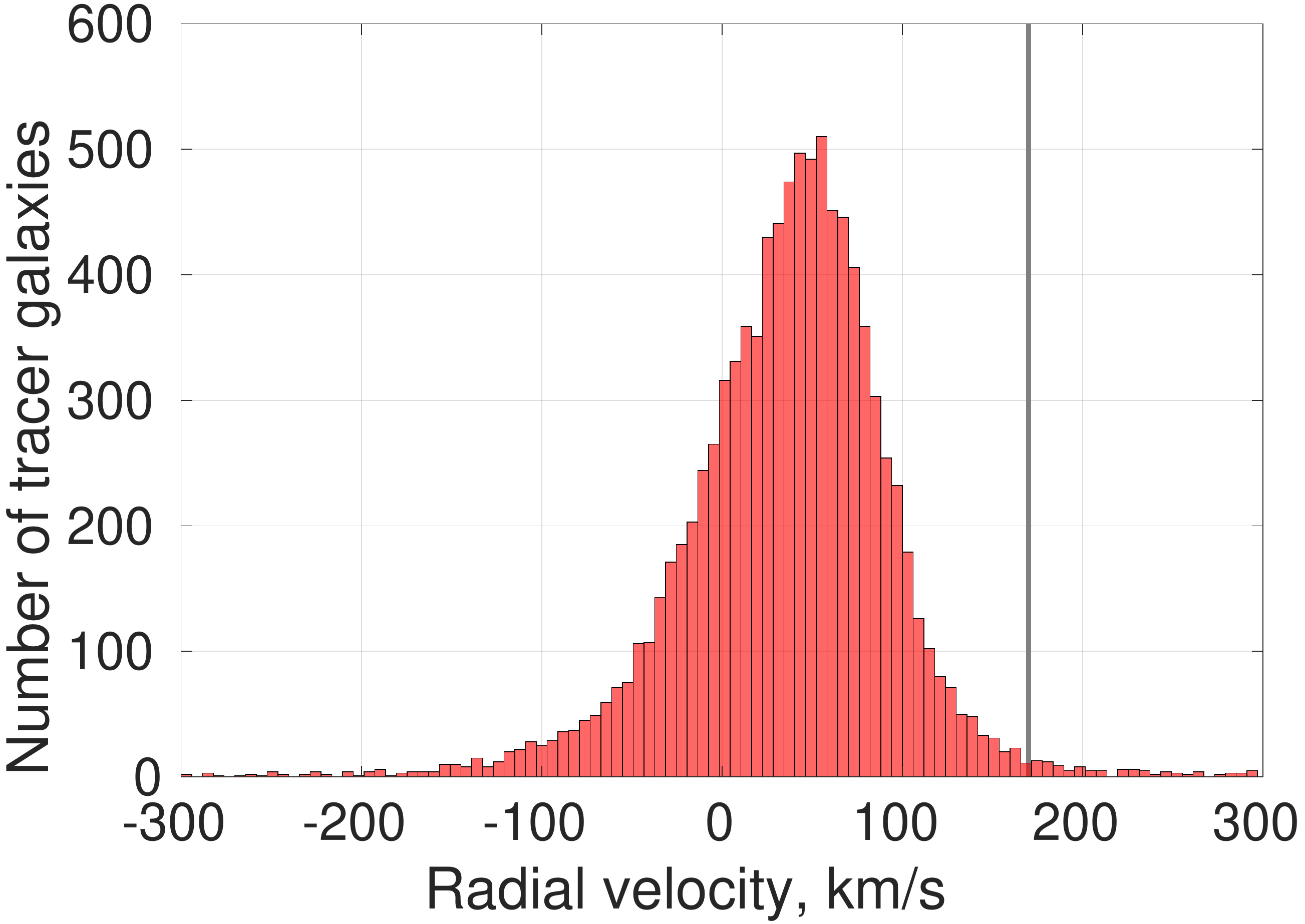}
	\caption{The RVs of tracer galaxies relative to their host within a distance range of $\left( 1.1 - 1.5 \right)$~Mpc, imposing the conditions in Section \ref{Observed_properties_NGC3109}. Additional restrictions are applied on LG-like hosts (see text). Results have been restricted to the RV range $\pm 300$~km/s for clarity. The RV of NGC~3109 is shown as a vertical grey line at 170~km/s \citep[figure 11 of][]{Banik_Zhao_2016}. 1.09\% of our tracers have a higher RV.}
	\label{v_r_all_hosts}
\end{figure}

The resulting distribution of RVs is shown in Figure \ref{v_r_all_hosts}, truncated to the range $\pm 300$~km/s for clarity. The observed Galactocentric RV of NGC~3109 is 170~km/s \citep[figure 11 of][]{Banik_Zhao_2016}, which we show using a vertical grey line. This lies above the vast majority of the distribution, which is consistent with figure~6 of \citet{Teyssier_2012} $-$ though their results were based on a much smaller sample size. Although we need to allow a wide enough distance range to build up the statistics, it is clear that the dispersion in RV is much larger than can be explained by variation of the Hubble flow velocity over the narrow distance range considered, thus demonstrating the power of a large cosmological simulation. Importantly, our results show that it is quite possible to have dwarf galaxies receding from the MW as fast as NGC~3109 at its observed distance and with a comparable mass. 1.09\% of our tracer galaxies have a higher RV, so the tension is mild. This decreases to 0.72\% if imposing isolation conditions on the NGC 3109 analogue as described in Section \ref{Backsplasher_isolation}.

\begin{figure}
	\centering
	\includegraphics[width = 8.5cm] {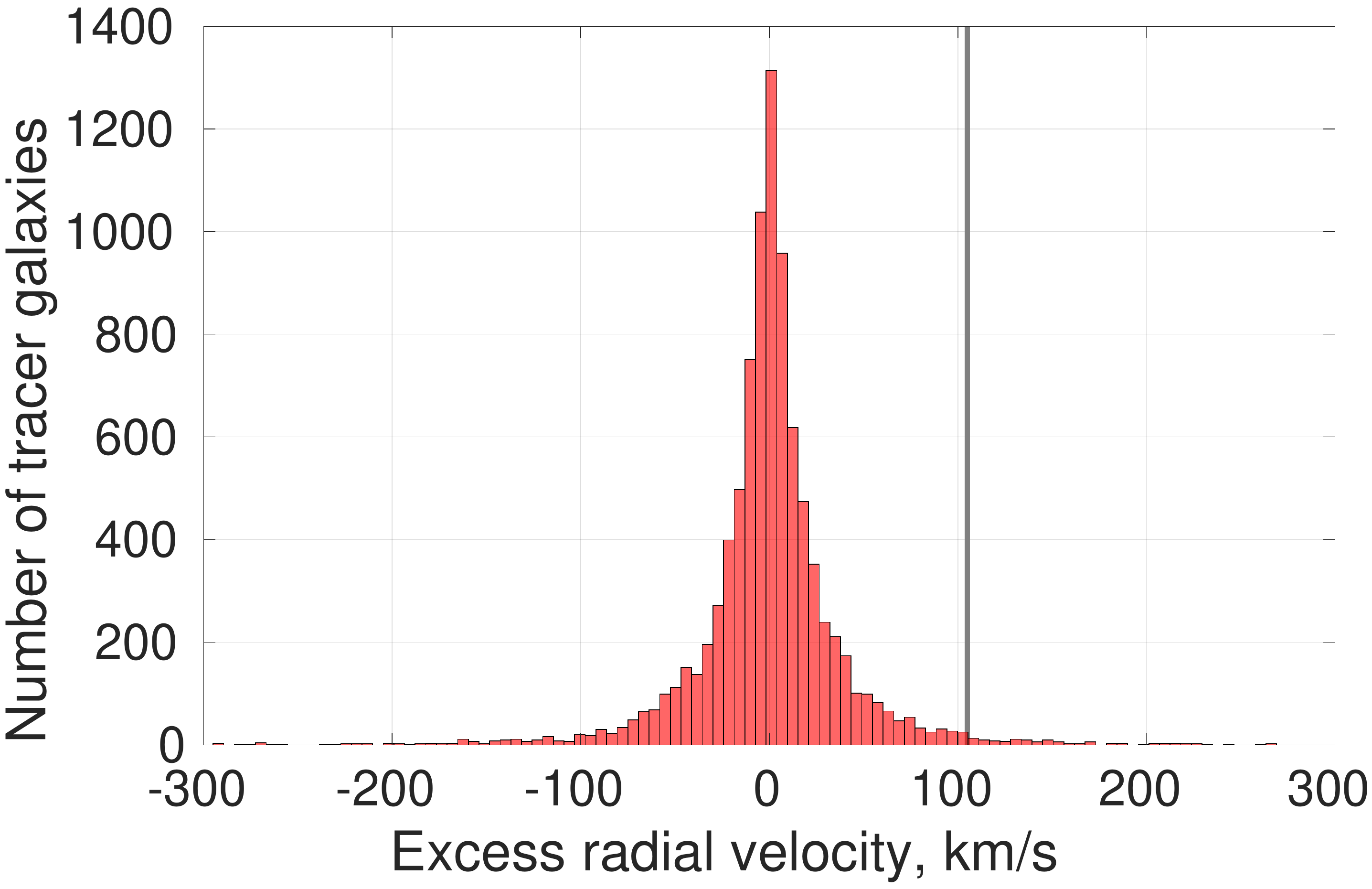}
	\caption{Similar to Figure \ref{v_r_all_hosts}, but after subtracting the mean RV of all tracer galaxies around the same host. We only show results for hosts with $\geq 5$ tracers. The vertical grey line at 105~km/s shows the RV excess of NGC~3109 relative to the best-fitting 3D $\Lambda$CDM timing argument analysis of \citet{Banik_2018_anisotropy}. 1.28\% of our tracers have a higher RV excess defined in the above sense. This is almost unchanged if restricting to only LG-like hosts (1.34\%; not shown). Note that the mean RV calculation for each host is allowed to take advantage of tracer galaxies with any mass, but only galaxies more massive than NGC~3109 are shown here (see text).}
	\label{Delta_v_r_all_hosts}
\end{figure}

Our results are based on stacking all host galaxies (with the above restriction on LG-like hosts). Since the hosts are not all equally massive, the expected RV at fixed distance will differ between hosts. To alleviate this, we next consider only hosts with $\geq 5$ tracer galaxies of any mass, allowing us to calculate their mass-weighted mean RV, and thus the distribution of tracer galaxy RVs around that mean. To improve the accuracy with which this mean is calculated, we relax the condition on the tracer galaxy's mass to compute the mean. We then subtract the mean RV from the tracer RV, and thereby determine the RV excess. Since not all hosts have 5 tracers with the appropriate position and at the same time at least one tracer more massive than NGC~3109, the statistics are somewhat noisier (Figure~\ref{Delta_v_r_all_hosts}). For NGC~3109, we show a vertical grey line at 105~km/s, its RV excess compared to the \citet{Banik_2018_anisotropy} timing argument analysis. In this case, 1.28\% of the distribution lies beyond 105~km/s. This decreases to 0.66\% if imposing isolation conditions on the NGC 3109 analogue as described in Section \ref{Backsplasher_isolation}. We conclude that ignoring the detailed observed environment of the LG, the 105 km/s RV excess of NGC~3109 is rare, but seemingly allowed at the percent level.

\subsection{Including the LG environment: timing argument calculations and their reliability}
\label{Timing_argument_reliability}

At a frequency of $\approx 1\%$, the RV of NGC~3109 is already uncommon, but not necessarily severely problematic if one neglects the environment around the LG. Nonetheless, considering the 3D positions of perturbers outside the LG in more detail should in principle account for much of the RV variation of the above-discussed analysis. This is precisely what was done in the 3D timing argument calculations of \citet{Banik_Zhao_2017}, \citet{Peebles_2017}, and \citet{Banik_2018_anisotropy}. A list of external perturbers taken into account can be found in table~3 of \citet{Banik_Zhao_2017}. Despite including all these perturbers and letting their masses vary, \citet{Peebles_2017} and \citet{Banik_2018_anisotropy} were nevertheless unable to account for the large observed RV of NGC~3109. It is therefore worth discussing whether their models can be trusted to accurately represent $\Lambda$CDM expectations for its RV.

Figure 9 of \citet{Banik_Zhao_2017} shows that if we suppose the model has a 25~km/s uncertainty, then it provides a good fit to the RVs of galaxies when the observed RV lies below the model prediction. The only exception is NGC 4163, but \citet{Peebles_2017} argued in his section 6.6 that it is part of the M94 group, and so excluded it from the timing argument analysis. As argued in section 5 of \citet{Banik_Zhao_2017}, NGC 4163 may well be a backsplasher flung towards us, as suggested by its RV being $\approx 100$~km/s lower than that of surrounding galaxies $-$ roughly the amount by which its RV falls below the model prediction. Excluding this problematic galaxy 3.0~Mpc away, a 25~km/s model uncertainty would nicely explain discrepancies between the model and observations in cases where the latter give a lower RV. It is also reasonable on theoretical grounds $-$ the dispersion in RV with respect to the LG barycentre at fixed distance from there should be $\approx 30$~km/s \citep{Aragon_Calvo_2011}. This suggests that even a 1D model of the LG should be accurate to about this much, so a 3D model can be expected to have an accuracy of $\approx 25$~km/s if not better \citep[see also section 4 of][]{Banik_Zhao_2016}.

Moreover, the timing argument is mostly sensitive to forces at late times, and thus mainly depends on the matter distribution today \citep[figure 4 of][]{Banik_Zhao_2016}. The model includes the Large Magellanic Cloud at a mass of $2.03 \times 10^{11} M_\odot$.\footnote{The reanalysis of \citet{Banik_2018_anisotropy} preferred a very slightly higher mass to that stated in table 3 of \citet{Banik_Zhao_2017}.} This automatically accounts for the induced reflex motion of the MW, which affects how we perceive the velocity field of the LG \citep{Jorge_2016}. 

Ideally, one should apply the same method to all our selected hosts and their environment in Illustris. While beyond the scope of the present paper, we note that when applied to 32 LG analogues in the Millennium simulation \citep{Springel_2005}, the method of \citet{Peebles_2011} gave reliable results for the total mass, i.e. the deviations between true and inferred masses were consistent with the inferred uncertainty \citep{Phelps_2013}. 

Therefore, with an expected accuracy of $\la 25$~km/s, it is not at all clear why the overall best-fitting model should underpredict the RV of a fairly massive isolated galaxy like NGC~3109 by $105 \pm 5$~km/s \citep[table 3 of][]{Banik_2018_anisotropy}. The discrepancy could thus be much more severe than the 1\% frequency we found in Section \ref{Hubble_diagram_analysis}. One explanation could be that some of the NGC~3109 analogues with a high RV are actually backsplashers, objects on rather extreme orbits that were once within the virial radius of their host but were subsequently carried outside of it. Indeed, one important aspect missing from the timing argument analyses is that they do not allow for dynamical friction on the extended dark matter haloes of galaxies, and the resulting mergers. They also do not account for the possibility of significant energy gain by a third galaxy near the spacetime location of the merger. To assess whether such backsplashers might resemble NGC~3109 today, we will in the rest of this paper investigate the distribution of backsplashers around our identified hosts in the Illustris TNG300 simulation.

\section{NGC 3109 as a backsplash galaxy}
\label{NGC3109abacksplash}

We now require that the analogues of NGC~3109 in the Illustris TNG300 simulation are backsplash galaxies.

\subsection{The backsplash condition}
\label{Backsplash_condition}

A backsplasher must have been within the virial radius of an MW analogue at some time in the past, but is by definition beyond its virial volume at the present time. To avoid making assumptions about the hypothesis being tested, we must allow for the possibility that the backsplasher is currently very far from the MW analogue that it interacted with. We keep the computational cost low by focusing on the virial volume of the MW analogue in all past snapshots for which $z \leq 5.22$, thus going slightly further back than \citet{Teyssier_2012}.

We use the Illustris merger tree \citep{Gomez_2015} to trace back the MW analogue subhalo identified at the present epoch. We find the virial radius $r_{vir}$ of the group to which it belongs at each timestep with $z \leq 5.22$, and then search through all the subhaloes within $r_{vir}$ of the subhalo corresponding to the MW analogue. A backsplasher candidate is defined as a subhalo in this list whose present-day root descendant lies beyond $2 \, r_{vir}$ from the MW analogue, with $r_{vir}$ being time-dependent. Some subhaloes within the virial volume at a past epoch are absent from the merger tree or have a root descendant at a previous epoch, i.e. they do not survive up to the present. We reject such cases from further consideration. To improve the efficiency of our algorithm, and since subhaloes within the virial volume of an MW analogue several Gyr ago may well have merged with it by now, we first check if the root descendant is the present MW analogue itself, and reject such cases.

It is quite possible that the same backsplasher is identified within the virial volume of an MW analogue at several past epochs. To avoid double-counting, we keep track of all subhaloes in the $z = 0$ subhalo catalogue. Once some subhalo $S$ in this catalogue has been recorded as a backsplash candidate, we ignore any subhalo at a previous epoch with root descendant $S$.

Since galaxies and subhaloes can merge in the hierarchical $\Lambda$CDM paradigm, we require that the {\it main progenitor} of a backsplasher was once within the virial radius of the host galaxy. Thus, each backsplash candidate is traced back along its main progenitor line to ensure that it was once within the virial radius of its host. In other words, we require that the bulk of the $z = 0$ subhalo has experienced a past backsplash encounter with the MW analogue. Without this restriction, we could have `backsplashers' which mostly consist of material that never passed within the virial radius of a massive galaxy. We thus avoid situations where a low-mass backsplasher subsequently merges with a nearby massive galaxy $A$, causing that the root descendant of the backsplasher is $A$. Such scenarios are not a viable explanation for NGC~3109 because it is highly unlikely for an even lower mass dwarf to be flung out at a very high speed, only to subsequently catch up and merge with NGC~3109. While the latter's RV would be somewhat affected, the scenario would not explain the anomalous kinematics of other galaxies in the NGC~3109 association. These are most likely not currently bound to NGC~3109 \citep{Kourkchi_2017} apart from Antlia, which is likely a satellite of NGC~3109 \citep[section 6.4 of][and references therein]{Banik_2018_anisotropy}. Thus, raising the RV of NGC~3109 alone would not be sufficient to raise the RVs of other association members. It also seems unlikely that a merger with NGC~3109 could raise its RV by 100~km/s without seriously disrupting its disc. %We checked that relaxing this condition does indeed lead to some very massive `backsplashers'.

\subsection{Isolation of the backsplasher}
\label{Backsplasher_isolation}

A particularly problematic aspect of NGC~3109 is its isolation, which implies that it was not substantially pulled away from the MW or M31 by a nearby massive group. Thus, NGC~3109 should have reached a Galactocentric distance of 1.2~Mpc without a significant `helping hand' from large scale structure. To avoid selecting dwarf galaxies in Illustris which did receive such a helping hand, we impose various isolation criteria on both the host galaxy and the NGC 3109 analogue. This is motivated by the observed distribution of matter in and around the LG. Table \ref{Perturber_distances} summarizes the locations of massive perturbers outside the LG, focusing on the distance from NGC~3109 and from the MW-M31 mid-point. While there are other dwarf galaxies at lower distances, these would have a very small effect on the trajectory of NGC~3109 \citep[e.g. see table 3 of][]{Banik_Zhao_2017}. Thus, we only impose isolation criteria in the sense of requiring no objects with sufficiently high mass and low separation.

\begin{table}
	\centering
	\begin{tabular}{ccc}
		\hline
		& \multicolumn{2}{c}{Distance in Mpc from} \\
		Perturber & MW-M31 mid-point & NGC 3109 \\ \hline
		Centaurus A & 4.19 & 3.06 \\
		M81 & 3.46 & 3.90 \\
		IC 342 & 3.14 & 4.11 \\
		\hline
	\end{tabular}
	\caption{Distances to massive objects outside the LG \citep[based on table 1 of][and references therein]{Banik_Zhao_2016}.}
	\label{Perturber_distances}
\end{table}

To get a similarly isolated object as NGC~3109, we require there to be no subhalo with $M > 5 \times 10^{11} M_\odot$ within 3~Mpc, with the mass threshold equal to the lowest allowed mass for an isolated host. An exception is made for the present-day descendant of the `host' subhalo that the backsplasher once interacted with.\footnote{The backsplasher and host need not be gravitationally bound either now or in the past, but the terminology is useful as the hypothesis being tested relates to the past configuration of the LG.} In cases where this host represents one member of an LG-like pair, we allow both members to be within 3~Mpc of the backsplasher. This imposes the condition that a 3~Mpc sphere centred on NGC~3109 contains no MW-mass galaxies other than the MW and M31, which is correct observationally (Table \ref{Perturber_distances}).

\subsection{Requiring energy gain}
\label{Requiring_energy_gain}

Our main purpose is to find trajectories with a similar final position to NGC~3109, but which would not be correctly modelled by the 3D timing argument analyses of \citet{Peebles_2017} and \citet{Banik_2018_anisotropy}. Merely passing through the virial radius of an MW-like host galaxy is not sufficient to invalidate especially the latter analysis, since it should have enough time resolution to correctly model the interaction $-$ partly because a softened force law was used to avoid singularities \citep{Banik_Zhao_2017}, leading to a smooth trajectory. Regardless of whether the MW mass distribution is modelled perfectly, a dwarf galaxy passing through would typically leave with the same energy as it came in because the model lacks dynamical friction. If such an \emph{energy-conserving} encounter also happens in Illustris, then we can be fairly confident that it would be appropriately modelled in the \citet{Peebles_2017} analysis and in that of \citet{Banik_2018_anisotropy}, which was very similar but had $10\times$ better temporal resolution.

During galaxy-galaxy encounters, dynamical friction plays an important role \citep[e.g.][]{Privon_2013, Kroupa_2015}. This would cause the backsplasher to lose energy, reducing its final RV and making it even more difficult to explain the anomalously high RV of NGC~3109. Therefore, only trajectories with significant energy gain might explain the anomalous kinematics of the HVGs. Such trajectories would very likely be mis-modelled in a few-body timing argument analysis, so they could represent a viable $\Lambda$CDM-based explanation.

To focus on such trajectories, we extract the host-backsplasher separation $d \left( t \right)$ using the merger tree (Section \ref{Simulation}). We define $v_{in}$ and $v_{out}$ as the backsplasher-host relative velocity at the times $t_{in}$ and $t_{out}$, respectively, when the backsplasher enters and leaves the region within $2 \, r_{vir, mid}$ of the host galaxy. We take $r_{vir, mid}$ to be the average of the virial radii at times $t_{in}$ and $t_{out}$, with a similar definition used for $M_{vir, mid}$ and the mid-point cosmic scale factor $a_{mid}$. We look backwards through the trajectory $d \left( t \right)$ until $d/r_{vir} < 2$, with $t_{out}$ being the snapshot when this first happens (looking backwards in time). $t_{out}$ is thus when the backsplasher crossed out of the twice-virial volume of its host, with $d_{out}$ being the separation at that time. To find a suitable choice for $t_{in}$, we look back even further to find the snapshot when $d/r_{vir}$ is lowest, which is approximately when the subhalo has its closest approach to the host. We stop looking back if $d/r_{vir} > 2$ again, ensuring that only the most recent encounter is considered in situations with multiple close encounters. We then consider all the snapshots $\leq 2$~Gyr prior to the point of closest approach (minimum $d/r_{vir}$), or back to the epoch when $a = 0.1$. Subject to these limits, we go backwards in time through the trajectory until $d > d_{out}$ again, choosing this to be our $t_{in}$. If this condition is not satisfied within our allowed time window, then we pick $t_{in}$ to be the snapshot in this time interval with the greatest $d$, thereby minimizing the gap with $d_{out}$ and allowing as fair a comparison as possible between $v_{in}$ and $v_{out}$. This scenario can arise if a backsplasher was originally a satellite orbiting well within the virial radius of its host since early times. Once we have found $t_{in}$, the separation at that time is defined as $d_{in}$.

The above procedure minimizes the difference between $d_{in}$ and $d_{out}$. Nonetheless, some difference remains, making it inaccurate to directly compare the relative velocities at those snapshots. We therefore define $\Delta v_{eff}$ as our measure of the energy gain, where
\begin{eqnarray}
	\Delta v_{eff} ~\equiv~ \sqrt{v_{out}^2 - v_{in}^2 + 2 \left( \Phi_{out} - \Phi_{in} \right)} \, ,
	\label{Delta_v_eff}
\end{eqnarray}
where $\Phi$ is the estimated specific potential energy of the backsplasher, with the subscripts indicating if the value is at time $t_{in}$ or $t_{out}$. To find $\Phi$, we assume that the host galaxy has a Navarro-Frenk-White density profile \citep{Navarro_1997} with the mass-concentration relation given in equation 4 of \citet{Duffy_2008}. Also adding an allowance for dark energy and dropping the in/out subscripts, we get that
\begin{eqnarray}
	\Phi ~&=&~ -\frac{G M_{vir, mid}}{d} \ln \left( 1 + \frac{d}{r_s} \right) - \frac{d^2{H_0}^2 \Omega_{\Lambda, 0}}{2} \, , \\
	r_s ~&\equiv&~ \frac{r_{vir, mid}}{c} \, , \\
	c ~&=&~ 6.71 \, {a_{mid}}^{0.44} \left( \frac{M_{vir, mid}}{M_0}\right)^{-0.091} \, ,
\end{eqnarray}
where the pivot mass $M_0 = 2 \times 10^{12} \, h^{-1} M_\odot$, and the present fraction of the cosmic critical density in dark energy is $\Omega_{\Lambda, 0} = 0.6911$. In cases where energy has been lost such that Equation \ref{Delta_v_eff} does not yield a real square root, we set
\begin{eqnarray}
	\Delta v_{eff} ~\equiv~ -\sqrt{-\left[ v_{out}^2 - v_{in}^2 + 2 \left( \Phi_{out} - \Phi_{in} \right) \right]} \, .
\end{eqnarray}
To minimize random fluctuations in our estimated potential, we use $M_{vir, mid}$ when calculating both $\Phi_{in}$ and $\Phi_{out}$, neglecting the possibility of a change in host mass over the period in which a backsplasher is inside the twice-virial volume. Although one can envisage more sophisticated schemes like considering the potential energy of each particle, this would involve handling a very large amount of particle-level data rather than the halo-level data used in our analyses, greatly increasing the computational cost. However, this would not much affect the results for a typical backsplash trajectory due to the good temporal resolution of the Illustris snapshots. Indeed, our results in Section \ref{Results} show that the potential adjustment term in Equation \ref{Delta_v_eff} is not very important. Moreover, our very conservative choice of threshold on $\Delta v_{eff}$ leaves a significant allowance for uncertainty in how it is calculated (see below).

\subsubsection{Toy model}

We now estimate the minimum $\Delta v_{eff}$ required to explain the anomalous RV of NGC~3109. For this purpose, we construct an idealized simulation in which a test particle moves under the influence of a galaxy with mass $M = 2 \times 10^{12} M_\odot$, which we refer to as the MW. As derived in section 2.1 of \citet{Banik_Zhao_2016} from General Relativity, the equation of motion for the particle position $\bm{r}$ relative to the galaxy contains a cosmological acceleration term in addition to the galaxy's gravity.
\begin{eqnarray}
	\ddot{\bm{r}} ~=~ \frac{\ddot{a}}{a} \bm{r} - \frac{GM \bm{r}}{\left(r^2 + r_c^2 \right) \sqrt{r^2 + r_s^2}} \, ,
	\label{Equation_of_motion}
\end{eqnarray}
where $r \equiv \left| \bm{r} \right|$, and an overdot denotes a time derivative. A force law of this form yields an extended region with a flat rotation curve of amplitude $v_f$, which fixes $r_s = GM/v_f^2$. We set $r_c = 0.01 \, r_s$ to prevent a singularity at the centre. To obtain an MW-like galaxy, we use $v_f = 180$~km/s \citep{Kafle_2012}. With these values, $r_s \approx r_{vir}$ at $z = 0$.

We start our fourth-order Runge-Kutta integration when $a = 0.1$, with the test particle having a peculiar velocity towards the galaxy in addition to some tangential velocity $v_{tan}$. Both parameters are varied to explore the parameter space. In each case, we must also choose the initial distance $d_i$ of the test particle, which sets its Hubble flow velocity. Our goal is to find trajectories which turn around and undergo a close encounter with the MW. At that time, the particle's speed $v$ is instantaneously increased as follows:
\begin{eqnarray}
	v ~\to~ \sqrt{v^2 + \left( \Delta v_{eff} \right)^2} \, .
	\label{Impulse}
\end{eqnarray}
This causes the particle to reach a larger distance than at first turnaround, which is typically at $\la 500$~kpc.

As $d_i$ is varied, perigalacticon occurs at different times. Increasing $d_i$ causes the particle to have more energy, which in turn causes it to encounter the MW later, and to leave with higher $v$. When $d_i$ is very small, increasing it significantly raises the post-encounter velocity while not much affecting the amount of time between the encounter and the present epoch, when our simulations end. This raises the present distance $d_f$. However, raising $d_i$ eventually causes the encounter to occur so late that $d_f$ starts decreasing again. We use a gradient ascent method \citep{Fletcher_1963} to maximize $d_f$ as a function of $d_i$.

\begin{figure}
	\centering
	\includegraphics[width = 8.5cm] {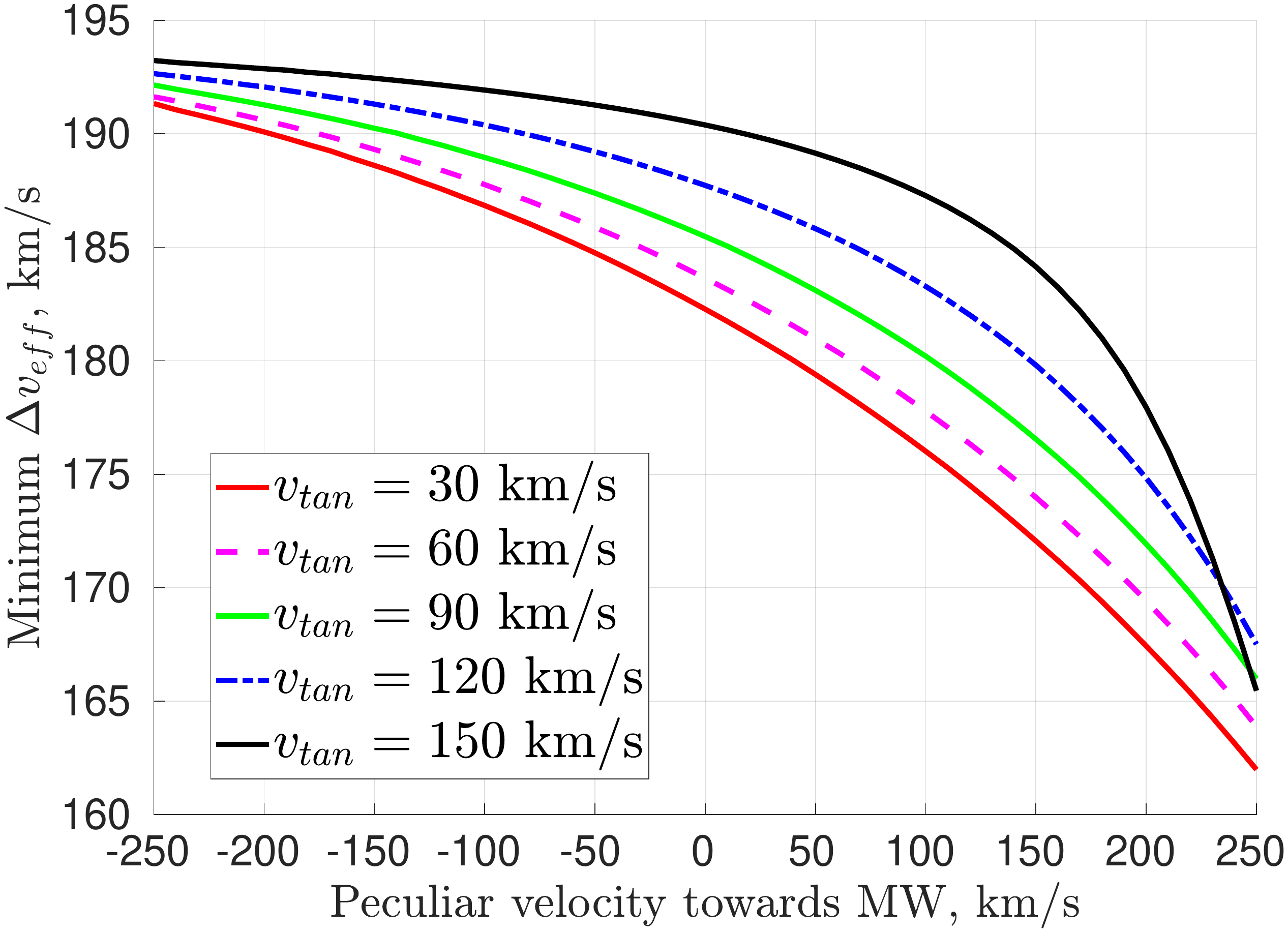}
	\caption{The minimum $\Delta v_{eff}$ (Equation \ref{Impulse}) required to reach a present distance of 1.2~Mpc after a previous close encounter with the MW. These results are based on idealized simulations that solve Equation \ref{Equation_of_motion}, which neglects (amongst other effects) dynamical friction and large-scale structure.}
	\label{Min_impulse}
\end{figure}

We repeat this procedure for a grid of initial radial and tangential peculiar velocities and $\Delta v_{eff}$. The range of $v_{tan}$ is limited above by the requirement to have a close encounter with the MW. At the lower limit, we expect results to depend very little on $v_{tan}$ as the orbit is essentially radial. We estimate that the radial component of the peculiar velocity has a $\pm 5\sigma$ uncertainty of $\pm 250$~km/s \citep[equation 16 of][]{Banik_Zhao_2017}. Within this range, we map out the minimum $\Delta v_{eff}$ that allows the particle to reach a post-encounter separation of 1.2~Mpc from the MW within a Hubble time. Our results are shown in Figure \ref{Min_impulse}. It is apparent that under conservative assumptions, we need $\Delta v_{eff} \ga 150$~km/s.

Assuming that such trajectories can be found in the Illustris simulation, we use our idealized setup to estimate the impact on the final RV. Without the impulse at pericentre, it is completely impossible for the particle to undergo a close approach to the MW and then reach a distance of 1.2~Mpc within the available timeframe. To facilitate a comparison, we construct a control non-impulsed ($\Delta v_{eff} = 0$) trajectory which never turns around and closely approaches the MW. This requires the use of a larger $d_i$.

An object ending up at larger $d_f$ generally has a larger RV, so this can be fairly compared between the trajectories only if they reach the same $d_f$. Thus, we vary $d_i$ for both the impulsed and the non-impulsed trajectories to ensure that $d_f = 1.2$~Mpc. With the impulsed trajectory, this implies the maximum possible $d_f > 1.2$~Mpc, so there are two possible choices for $d_i$. We choose the larger $d_i$ since this causes the perigalacticon to occur later, implying the particle must have a larger final RV to reach the same $d_f$. This lets us find how much the final RV could differ between the impulsed and control trajectories. In both cases, we use a Newton-Raphson algorithm to vary $d_i$ in order to precisely achieve $d_f = 1.2$~Mpc.

\begin{figure}
	\centering
	\includegraphics[width = 8.5cm] {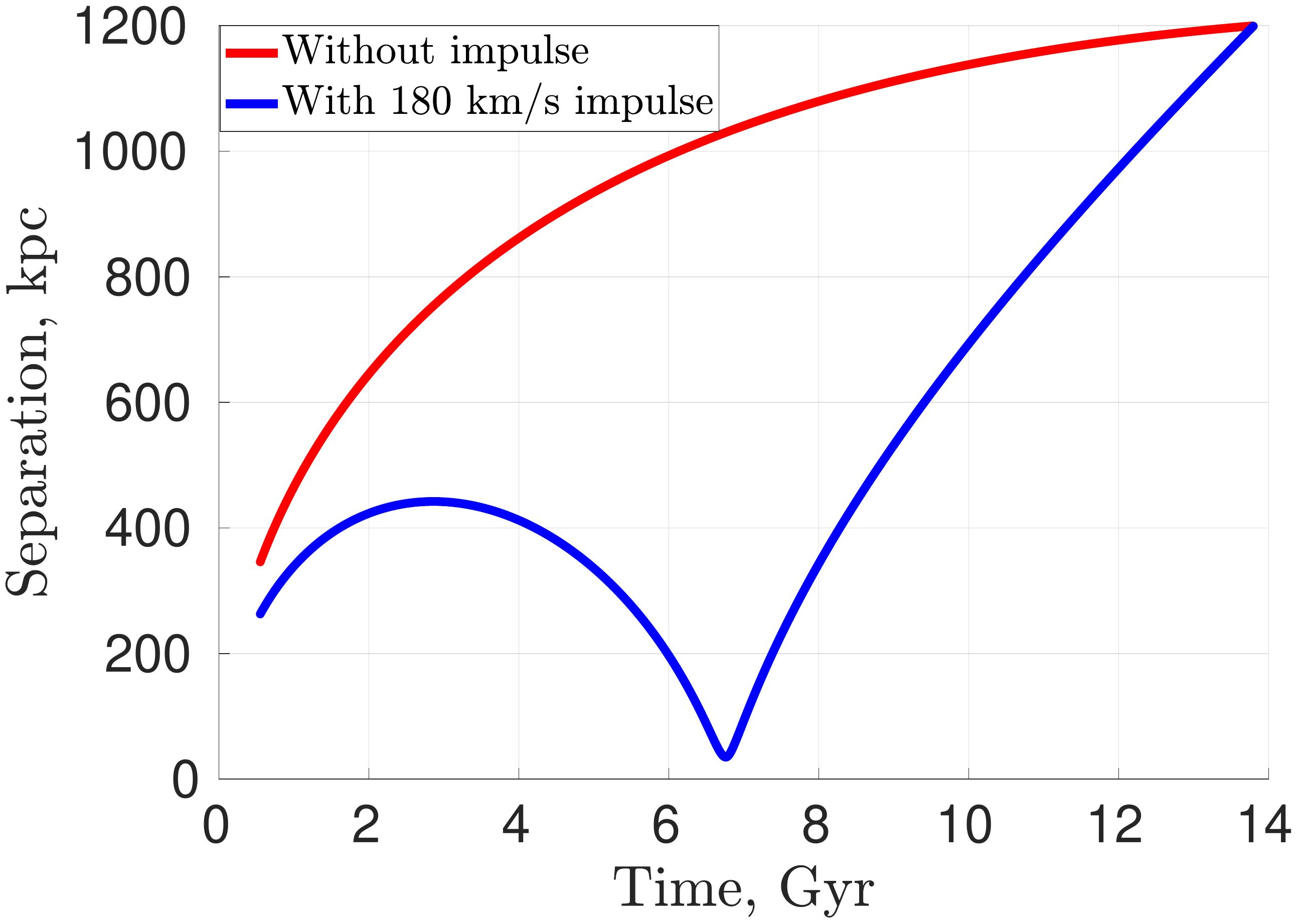}
	\caption{Example of a trajectory (blue) which satisfies Equation \ref{Equation_of_motion} with $\Delta v_{eff} = 180$~km/s, $v_{tan} = 50$~km/s, and a radial peculiar velocity towards the MW of 100~km/s when $a = 0.1$. For comparison, we show a similar trajectory with larger initial separation and no close encounter with the MW such that $\Delta v_{eff} = 0$ (red). Both trajectories reach the same present distance of 1.2~Mpc, but the present RV of the impulsed trajectory is 109.9~km/s higher. Another impulsed trajectory can also be constructed with the same initial peculiar velocity and final distance but with lower $d_i$, an earlier encounter, and lower final RV (not shown).}
	\label{Example_trajectory_3_17_36}
\end{figure}

Figure \ref{Example_trajectory_3_17_36} shows an example where the impulsed trajectory has $\Delta v_{eff} = 180$~km/s, $v_{tan} = 50$~km/s, and an initial radial peculiar velocity towards the MW of 100~km/s. For the non-impulsed trajectory, we use the same initial peculiar velocity but larger $d_i$. The final RV is 9.73~km/s for the non-impulsed trajectory and 119.64~km/s for the impulsed trajectory. It is clear that the impulse has provided an alternative high-velocity route to reaching the presently observed distance of NGC~3109. The RV excess of this route compared to the `traditional' (non-impulsed) route is 110~km/s. This would nicely explain why the RV of NGC~3109 exceeds the prediction of the \citet{Banik_2018_anisotropy} model by 105~km/s.

We next consider whether backsplashers in the Illustris TNG300 cosmological simulation with the mass and present distance of NGC~3109 ever have trajectories with a similarly large $\Delta v_{eff}$. This is possible if a dwarf galaxy closely encounters the MW while it is undergoing a minor merger $-$ a significant amount of energy could be gained in a three-body interaction. But such scenarios could prove too rare, or dynamical friction on the dwarf could slow it down such that there is a net loss of orbital energy ($\Delta v_{eff} < 0$). Our simplified analysis in this section neglected the role of dynamical friction, which could be important at the mass of NGC~3109 (Section \ref{Discussion}).

%Preparing trajectories for vtan = 50 km/s, impulse = 200 km/s, vpec towards MW = 100 km/s.
%Impulse = 0 km/s, final RV is vr = 21.3305 km/s.
%Impulse = 200 km/s, final RV is vr = 150.7762 km/s.
%However, we argue in the text that such trajectories do not arise in self-consistent $\Lambda$CDM simulations for objects as massive as NGC 3109. Dynamical friction must be at least partly responsible.

\section{Results}
\label{Results}

%The RV of NGC 3109 is apparently not problematic if one neglects the environment around the LG. Nonetheless, considering the 3D positions of perturbers outside the LG in more detail should in principle account for much of the above-mentioned RV variation. This is precisely what was done in the 3D timing argument calculations of \citet{Peebles_2017} and \citet{Banik_2018_anisotropy}. However, these analyses do not allow for mergers between galaxies, and thus the possibility of significant energy gain by a third galaxy near the spacetime location of the merger. To assess whether such backsplashers might resemble NGC 3109 today.

\begin{figure}
	\centering
	\includegraphics[width = 8.5cm] {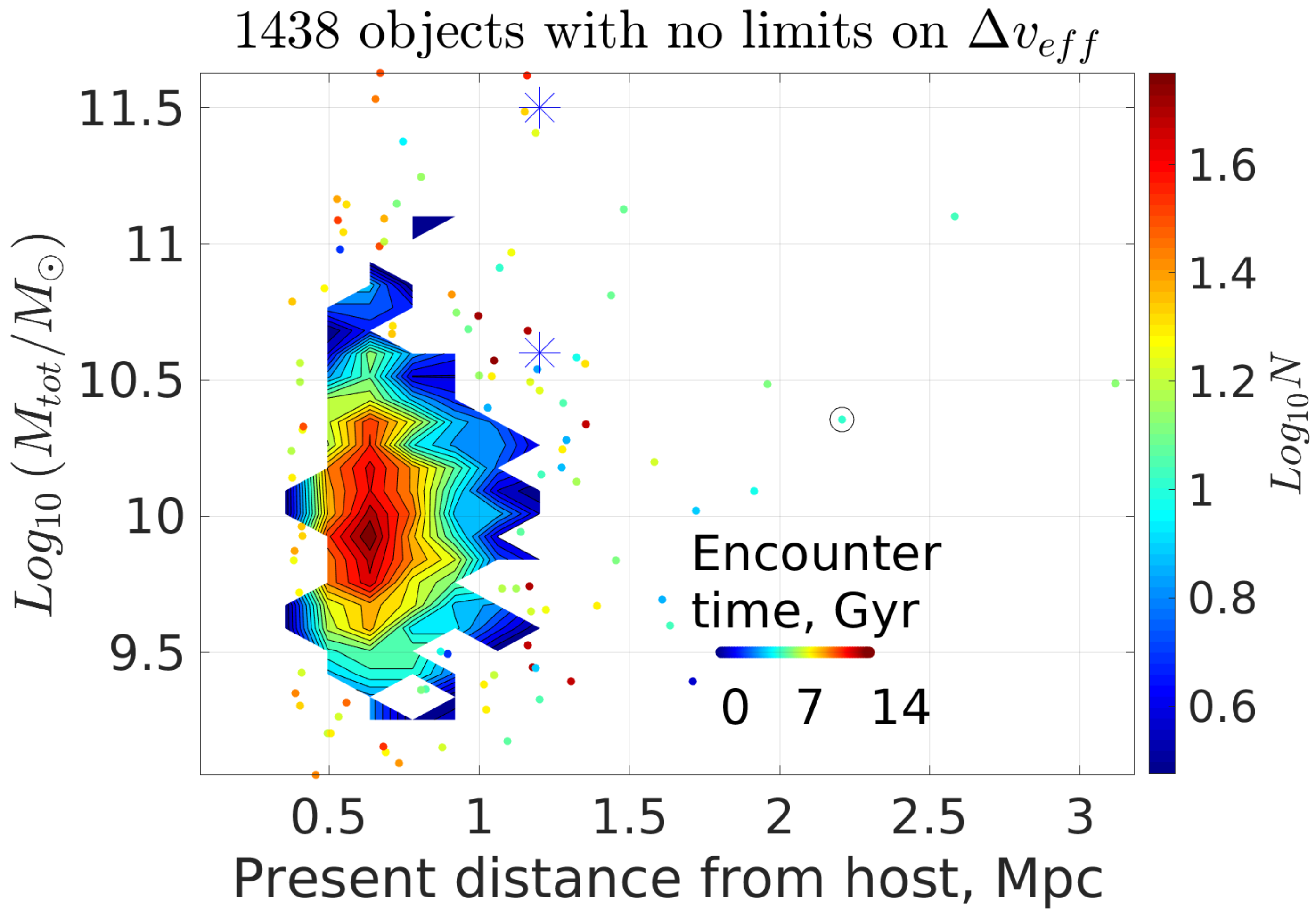}
	\caption{Distribution of total mass and distance from host for the backsplashers we identify in the Illustris TNG300 simulation. No restriction is imposed on the energy gain $\Delta v_{eff}$ during the encounter. Outside the high-density region (contours), we show individual backsplashers, with the colour indicating the encounter time when $d/r_{vir}$ was lowest (Section \ref{Requiring_energy_gain}). The distance and virial mass of NGC~3109 are shown as blue stars for two possible masses (Table \ref{NGC_3109_parameters}). The total number of backsplashers is indicated at the top. The trajectory of the circled backsplasher slightly right of centre is shown in solid black in Figure \ref{Illustris_trajectories}.}
	\label{M_vir_Delta_v_no_limits}
\end{figure}

We begin by showing the distribution of backsplasher total mass and present distance from the host (Figure \ref{M_vir_Delta_v_no_limits}). In case of an LG-like host, we show the minimum distance from either of the host galaxies. For now, we do not impose any restriction on $\Delta v_{eff}$. Without this restriction, we identify 1438 backsplashers. The vast majority of these lie at distances ${\la 1}$~Mpc and mass ${\la 10^{10.5} M_\odot}$. Both the distance and mass of NGC~3109 are individually highly unlikely if it is drawn from the distribution of backsplashers in $\Lambda$CDM. Only a handful of backsplashers match NGC~3109 in both respects, and even then only if we use the lower possible NGC~3109 mass of ${10^{10.6} M_\odot}$. This is only just allowed in some Newtonian rotation curve fits (Figure \ref{Li_results}).

\begin{figure}
	\centering
	\includegraphics[width = 8.5cm] {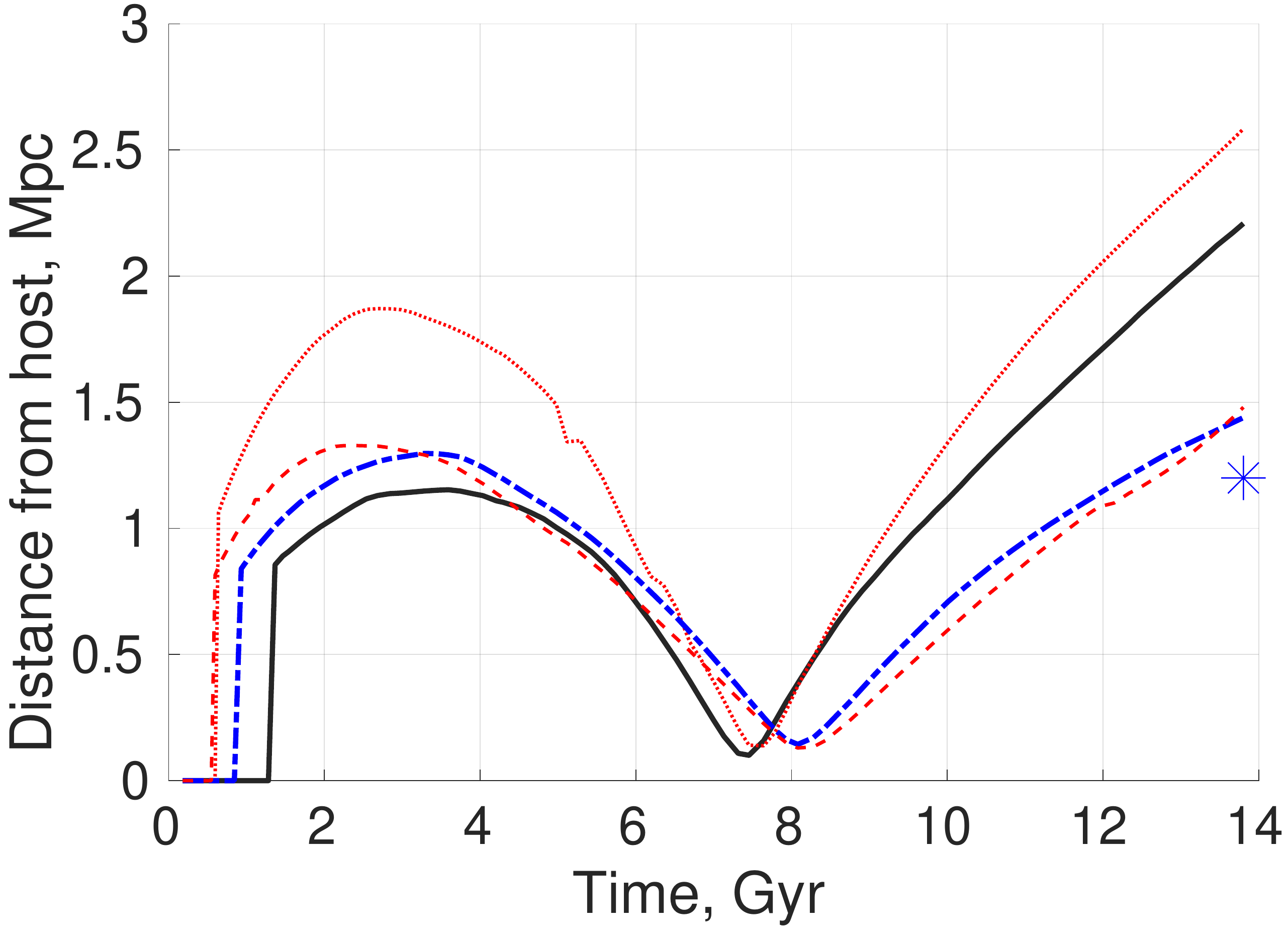}
	\includegraphics[width = 8.5cm] {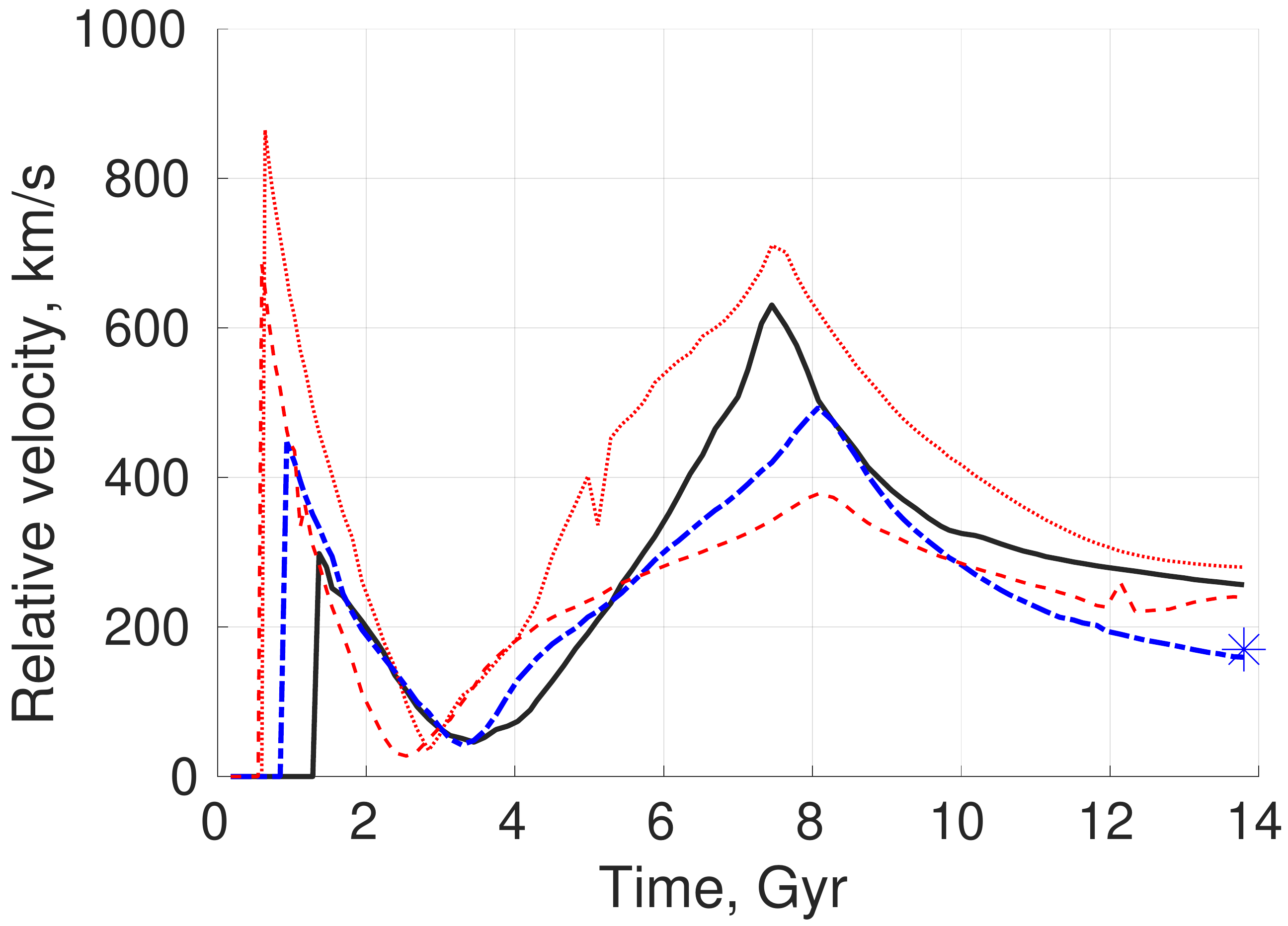}
	\caption{Host-backsplasher separation (\underline{top}) and relative velocity (\underline{bottom}) for the three subhaloes with larger mass and host-centric distance than NGC~3109 (Figure \ref{M_vir_Delta_v_no_limits}). The blue star in the top panel indicates the present position of NGC~3109 (Section \ref{NGC3109_distance}), while the blue star in the bottom panel indicates its Galactocentric RV. Notice that the red and blue trajectories appear symmetric around pericentre, and thus show little sign of energy gain while interacting with their host. This is borne out by their somewhat negative $\Delta v_{eff}$ (see text). For comparison, we also show the trajectory of the most distant backsplasher in our sample with $\Delta v_{eff} \geq 0$ (solid black), even though its mass is less than that of NGC~3109 (circled object in Figure \ref{M_vir_Delta_v_no_limits}). In this case, energy gain is apparent ($\Delta v_{eff} = 225$~km/s).}
	\label{Illustris_trajectories}
\end{figure}

Our results in Figure \ref{M_vir_Delta_v_no_limits} suggest that it is sometimes possible to get backsplashers with the mass and distance of NGC~3109. However, we have not yet considered whether the trajectories of these three backsplashers truly represent behaviour that would not be captured by the modelling of \citet{Banik_2018_anisotropy}. To investigate this further, we use Figure \ref{Illustris_trajectories} to show the distance and relative velocity between these backsplashers and their hosts as a function of time. It is evident that in all three cases (curves except that in solid black), the trajectory appears symmetric around the time of closest approach to the host. This is borne out by the values of $\Delta v_{eff}$ in km/s with (without) the potential energy adjustment term in Equation \ref{Delta_v_eff}, which in increasing order of final distance are $-160$ ($-151$), $-90$ ($-90$), and $-121$ ($-104$), indicating energy loss in all cases. For illustrative purposes, we also show a fourth backsplasher (solid black) with a lower mass of $2.26 \times 10^{10} M_\odot$. In this case, the backsplasher has clearly gained energy during the encounter, as also evident in that its $\Delta v_{eff} = 225$~km/s. The good time resolution of the Illustris snapshots allows us to measure the relative velocity at essentially the same separation from the host before and after the encounter, minimizing the potential energy adjustment (if we neglect this, we get $\Delta v_{eff} = 247$~km/s). Thus, the Illustris TNG300 simulation contains genuine backsplash trajectories with $\Delta v_{eff} > 0$, but not at the high mass and distance of NGC~3109.

Our simplified model in Section \ref{Requiring_energy_gain} suggests that trajectories with $\Delta v_{eff} \la 150$~km/s are unable to reach a present distance of 1.2~Mpc. In a cosmological simulation, the effects of large scale structure allow a dwarf to reach this distance despite having a close encounter with an MW analogue which yields zero or even negative $\Delta v_{eff}$. However, as discussed in Section \ref{Requiring_energy_gain}, such trajectories should be modelled correctly in the 3D timing argument analyses of \citet{Peebles_2017} and \citet{Banik_2018_anisotropy}. Thus, the failure of their model to correctly represent NGC~3109 cannot be understood using Illustris trajectories with $\Delta v_{eff} < 0$. If anything, loss of energy during a past close interaction with the MW (e.g. due to dynamical friction) would make it even more difficult to explain the anomalously high RV of NGC~3109.

\begin{figure}
	\centering
	\includegraphics[width = 8.5cm] {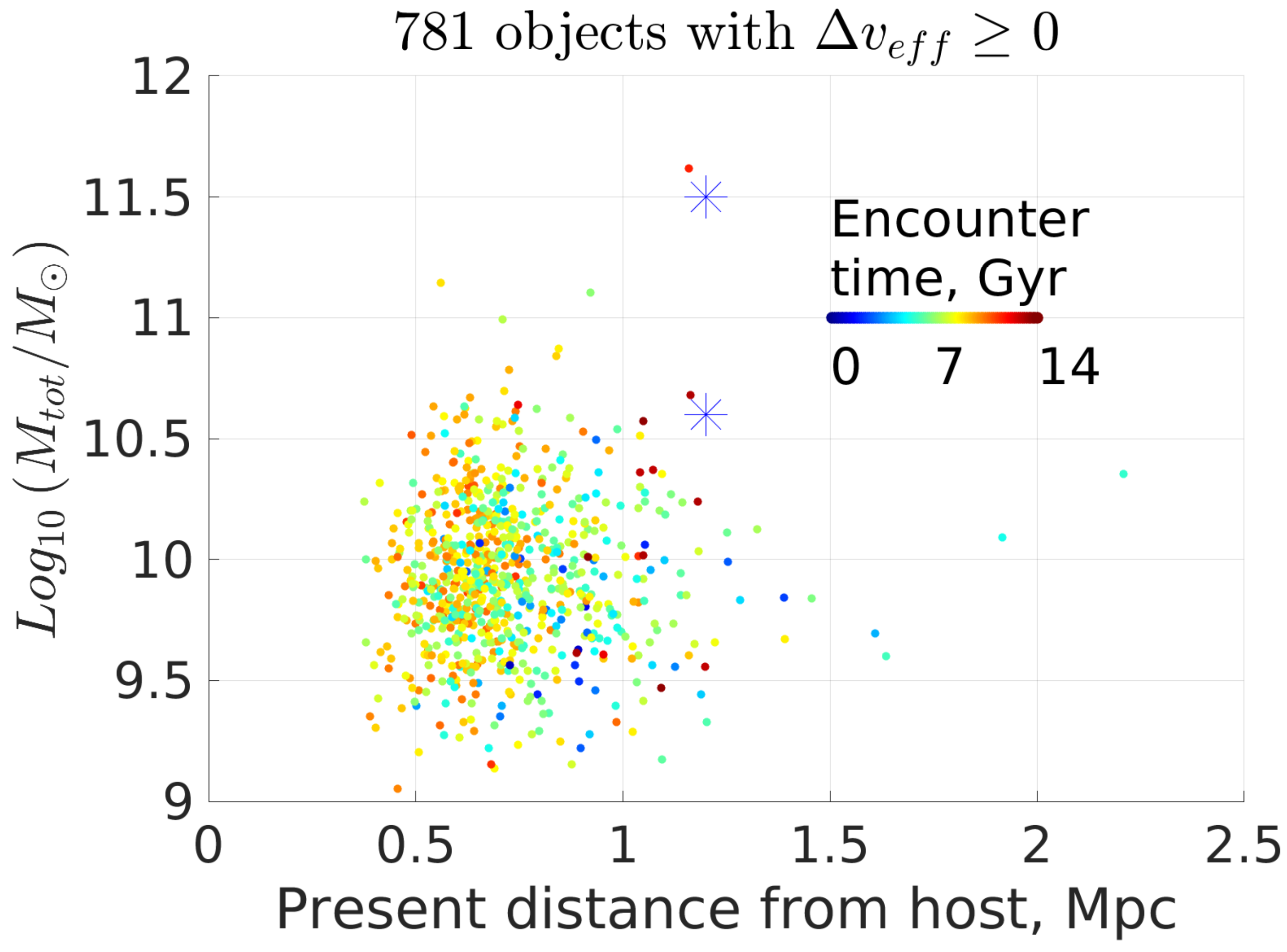}
	\caption{Similar to Figure \ref{M_vir_Delta_v_no_limits}, but now requiring $\Delta v_{eff} \geq 0$. This removes all backsplashers with greater mass and host-centric distance than NGC~3109. We show all backsplashers individually as there are not enough to reliably draw contours.}
	\label{M_vir_Delta_v_0}
\end{figure}

To be very conservative, we first consider the distribution of backsplashers if we merely require that $\Delta v_{eff} \geq 0$. Our results show no backsplash analogues to NGC~3109 (Figure \ref{M_vir_Delta_v_0}). We can find one analogue if we reduce the required mass to $2.26 \times 10^{10} M_\odot$. However, the rotation curve of NGC~3109 reaches an amplitude of $\approx 80$~km/s at a distance of 12~kpc, and is likely still rising there \citep[figure 13 of][]{Carignan_2013}. This implies a Newtonian dynamical mass of $1.8 \times 10^{10} M_\odot$ within 12~kpc, making it highly unlikely that the virial mass of NGC~3109 is only $2.3 \times 10^{10} M_\odot$. Plausible rotation curve fits in a $\Lambda$CDM context yield significantly larger values, with none suggesting a mass below $10^{10.6} M_\odot = 4.0 \times 10^{10} M_\odot$ (Figure \ref{Li_results}).

\begin{figure}
	\centering
	\includegraphics[width = 8.5cm] {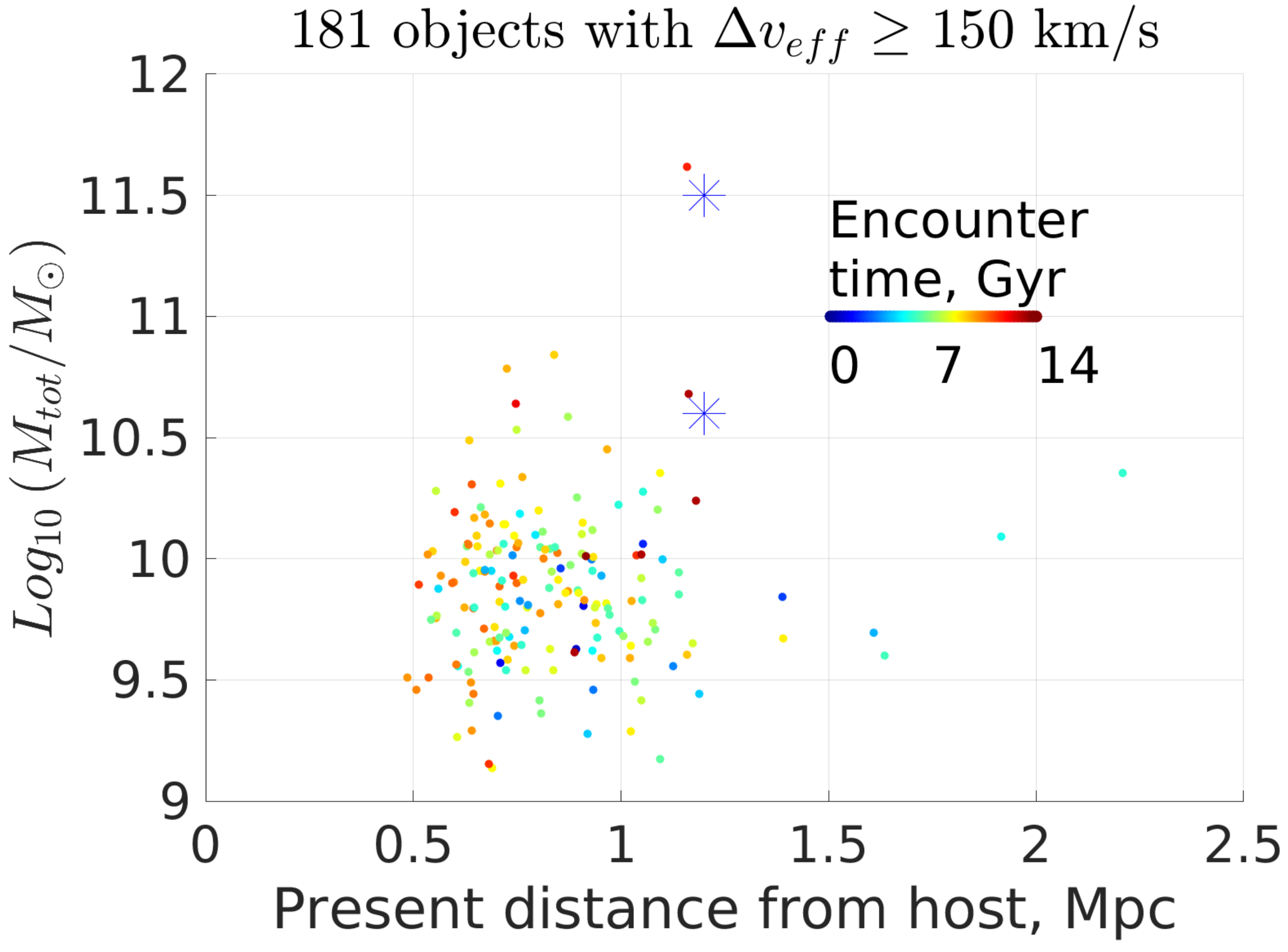}
	\caption{Similar to Figure \ref{M_vir_Delta_v_no_limits}, but now requiring $\Delta v_{eff} \geq 150$~km/s. We argued in Section \ref{Requiring_energy_gain} that this is required to explain the anomalous kinematics of NGC~3109.}
	\label{M_vir_Delta_v_150}
\end{figure}

Our results in Figure \ref{Min_impulse} suggest that $\Delta v_{eff} \ga 150$~km/s for a backsplasher to reach the present distance of NGC~3109 along a substantially different route (and hence different final RV) to a non-backsplash galaxy at the same present position. A smaller impulse would mean the object had more of a helping hand from large scale structure, weakening the case that its trajectory would not be correctly modelled by \citet{Banik_2018_anisotropy}. Thus, we use Figure \ref{M_vir_Delta_v_150} to show the effect of requiring $\Delta v_{eff} \geq 150$~km/s. The overall distribution of backsplashers remains similar, but their number is reduced more than four-fold. NGC~3109 is now much further from the backsplashers' distance and mass distribution. This is especially true if we assign NGC~3109 a mass of $10^{11.5} M_\odot$, as would be required to once have bound the whole NGC~3109 association \citep[section 3 of][]{Bellazzini_2013}. We discuss this issue further in Section \ref{Discussion}.

\begin{figure}
	\centering
	\includegraphics[width = 8.5cm] {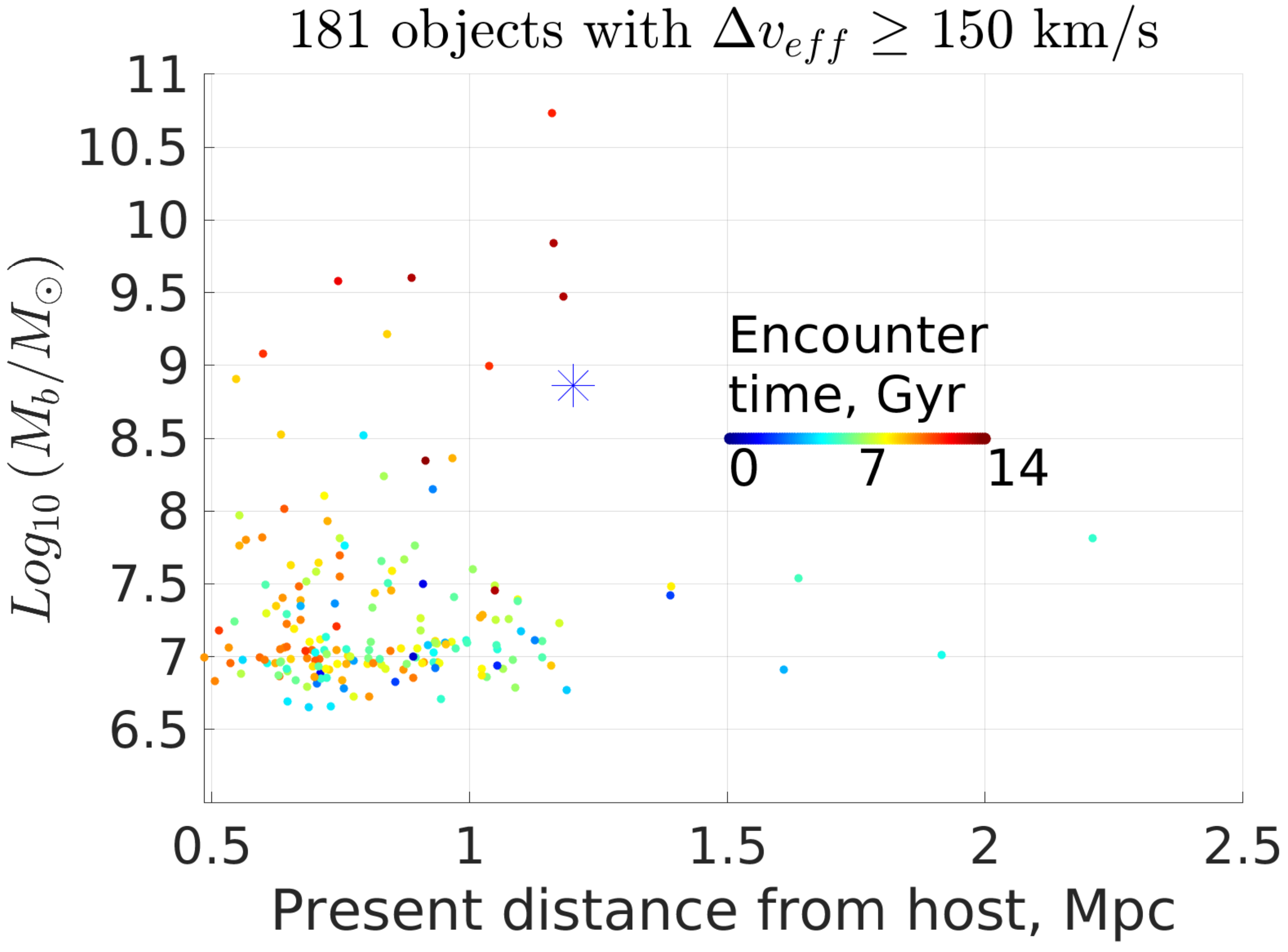}
	\includegraphics[width = 8.5cm] {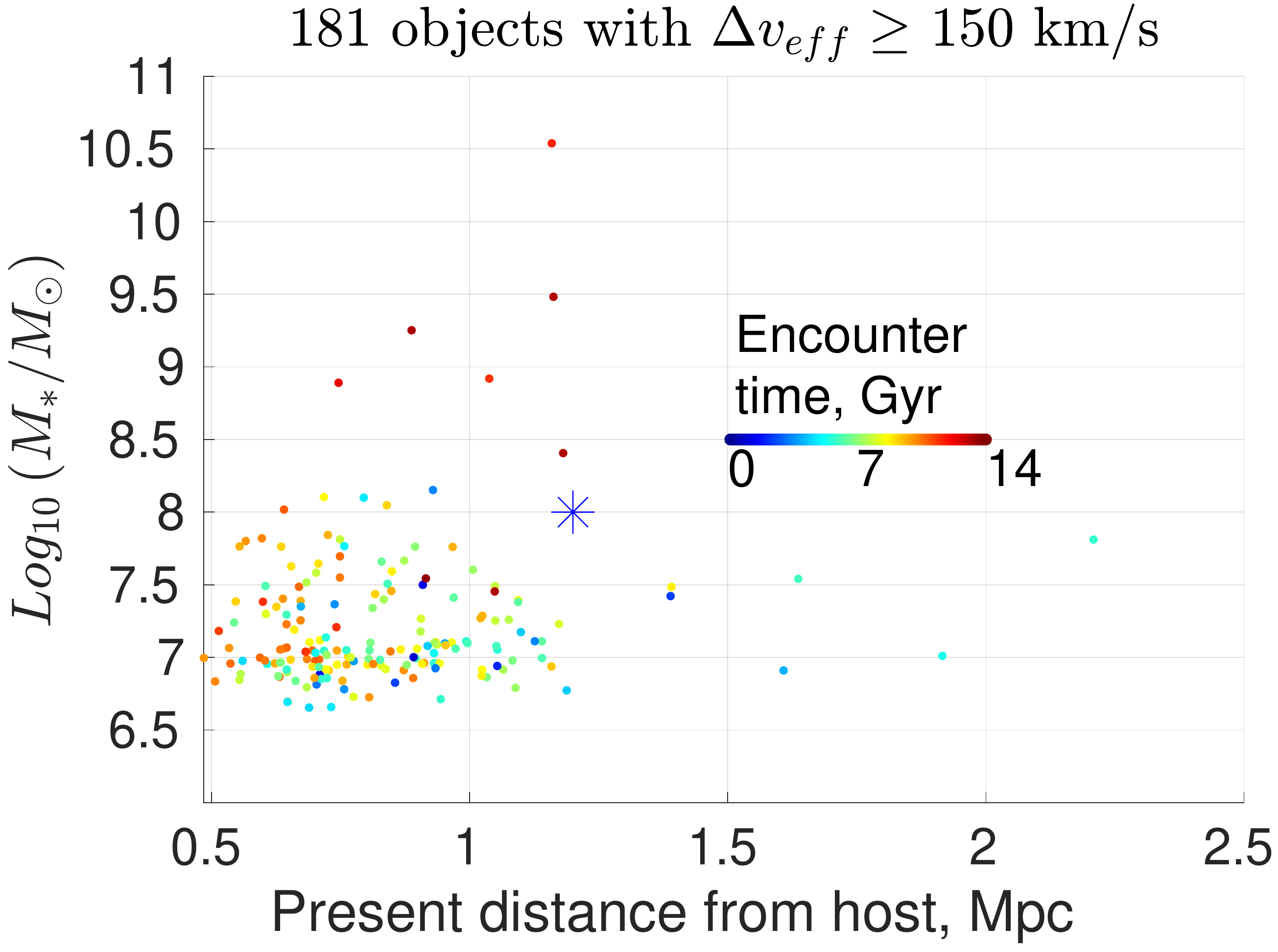}
	\caption{Similar to Figure \ref{M_vir_Delta_v_150}, but now showing only the baryonic mass (\emph{top}) or stellar mass (\emph{bottom}). In each panel, the corresponding observables for NGC~3109 are shown as a blue star (Table~\ref{NGC_3109_parameters}).}
	\label{M_b_Delta_v_150}
\end{figure}

Since Illustris is a hydrodynamical simulation, we can also consider the baryonic and stellar masses of the backsplashers we have identified. The results confirm that the distance of NGC~3109 and its baryonic or stellar mass are indeed significantly higher than expected for backsplashers in $\Lambda$CDM (Figure \ref{M_b_Delta_v_150}). In particular, the bottom panel shows that the well-constrained stellar mass of NGC~3109 is quite unusual for a backsplasher at any distance. This could be related to ram pressure stripping of the backsplasher's gas while it closely encounters a more massive galaxy \citep[see also][]{Teyssier_2012}.

\begin{figure}
	\centering
	\includegraphics[width = 8.5cm] {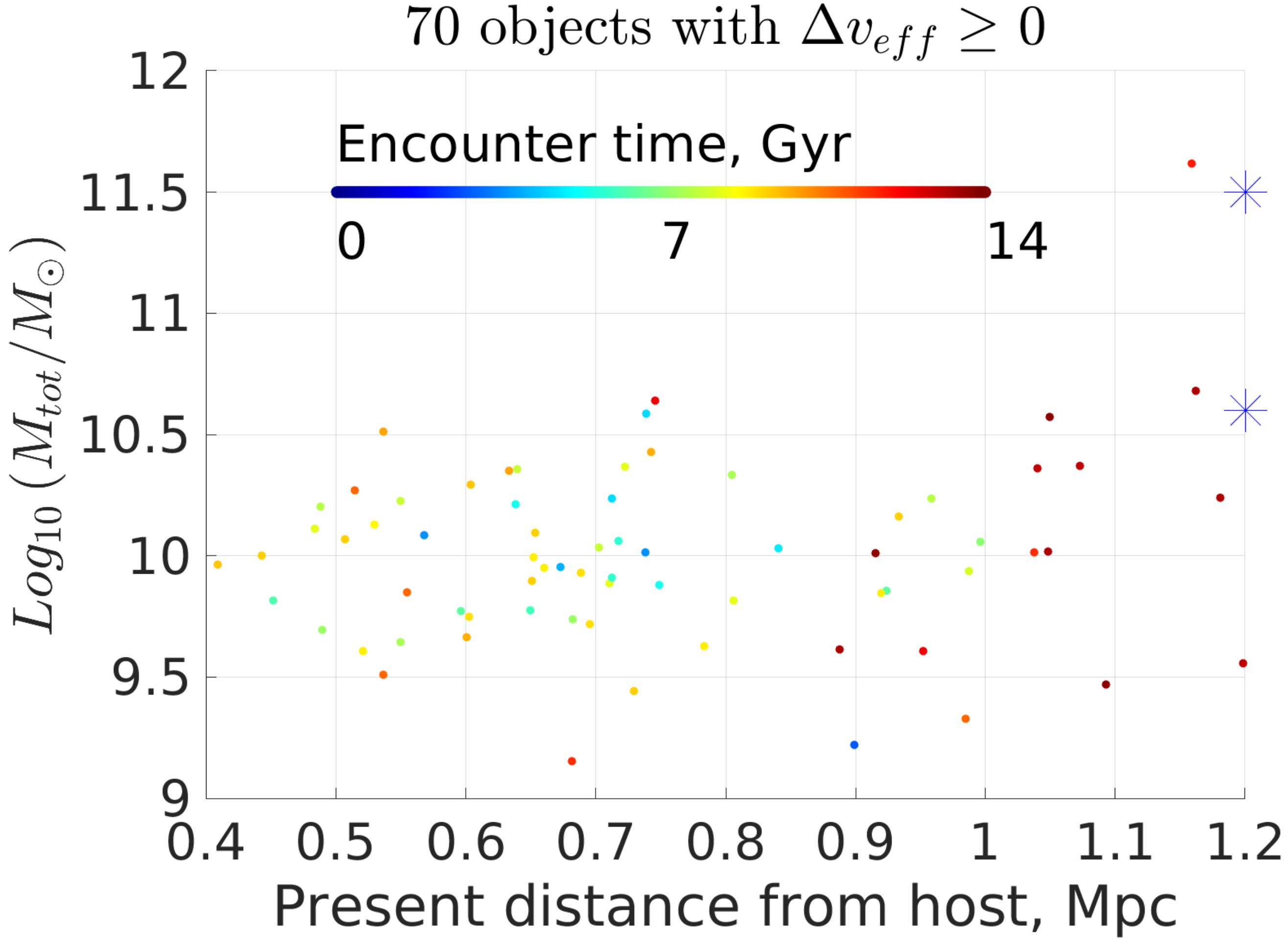}
	\caption{Similar to Figure \ref{M_vir_Delta_v_0}, but considering only LG-like hosts (see text). Notice the narrower range of distance, causing the properties of NGC~3109 to appear at the right edge (blue stars).}
	\label{M_vir_Delta_v_0_LGlike_only}
\end{figure}

So far, we have not distinguished between whether the host galaxy of a backsplasher is isolated or in an LG-like pair. This allows us to build up much better statistics, since we only have 640 host galaxies in 320 LG-like paired configurations. However, this is sufficient to get a good idea if the mass-distance distribution of backsplashers is similar around LG-like host galaxies. We therefore conduct a similar analysis to Figure \ref{M_vir_Delta_v_0} but only for LG-like hosts, with the result shown in Figure \ref{M_vir_Delta_v_0_LGlike_only}. It is clear that the overall distributions are very similar, though the smaller number of objects in the latter case causes the tail to be sampled less well. As a result, there are now no backsplashers as distant as NGC~3109 for any mass, even with our very conservative distance estimate of 1.2~Mpc. The similarity of results between LG-like hosts and the full sample (with mostly isolated hosts) indicates that the presence of M31 does not make it easier to explain how NGC~3109 could be a backsplasher in a conventional gravity context.

Our results allow us to consider whether NGC~3109 could be a backsplasher from M31 rather than the MW. This would require a present distance from the host of 2.0~Mpc (Section \ref{NGC3109_distance}). However, none of the $\Delta v_{eff} \geq 0$ backsplashers associated with LG-like hosts reach a separation of even 1.2~Mpc, and generally also have a much lower mass than NGC~3109. Clearly, a 2~Mpc separation with the host would make NGC~3109 significantly more of an outlier from the expected backsplasher distribution of distance and mass. Therefore, the (very small) probability of NGC~3109 being a $\Lambda$CDM backsplasher arises mostly from the chance that a suitable backsplash event occurred near the MW.

\subsection{Comparison with dark matter-only simulation}
\label{Dark_matter_only}

%12960 hosts, 546 are in 273 LG-like pairs.
%Level of significance is 3.9529 standard deviations.
%Hello, analogue found with log10(M) = 10.7592, d = 1.2605, Delta_v_eff = 105.2399 km/s, potential adjustment = -62.748 km/s, Delta_v_eff(fixed Phi) = 122.5265 km/s, t_min/Gyr = 4.8159.
%Hello, analogue found with log10(M) = 10.7592, d = 1.2605, t_entry/Gyr = 4.0375, t_exit/Gyr = 5.5766.
%Hello, analogue found with log10(M) = 10.7592, r_vir_entry/kpc = 93.8893, r_vir_exit/kpc = 143.5877.
%Hello, analogue found with log10(M) = 10.7592, M_vir_entry/M_Sun = 5.427097e+11, M_vir_exit/M_Sun = 1.088978e+12.
%Hello, analogue found with log10(M) = 10.7592, M_backsplasher_entry/M_Sun = 5.205882e+10, M_backsplasher_exit/M_Sun = 4.745308e+10.
%Backsplash_subhaloID = 3979932.
%Backsplash_host = 8749.

Our analysis thus far has focused exclusively on the Illustris TNG300 simulation. This can easily resolve haloes with the mass of NGC~3109 (Figure \ref{M_vir_Delta_v_no_limits}), while the larger simulation volume than e.g. TNG100 should allow for better statistics. The backsplash process mainly revolves around the motions of fairly massive galaxies, so baryonic physics should play only a small role.

\begin{figure}
	\centering
	\includegraphics[width = 8.5cm] {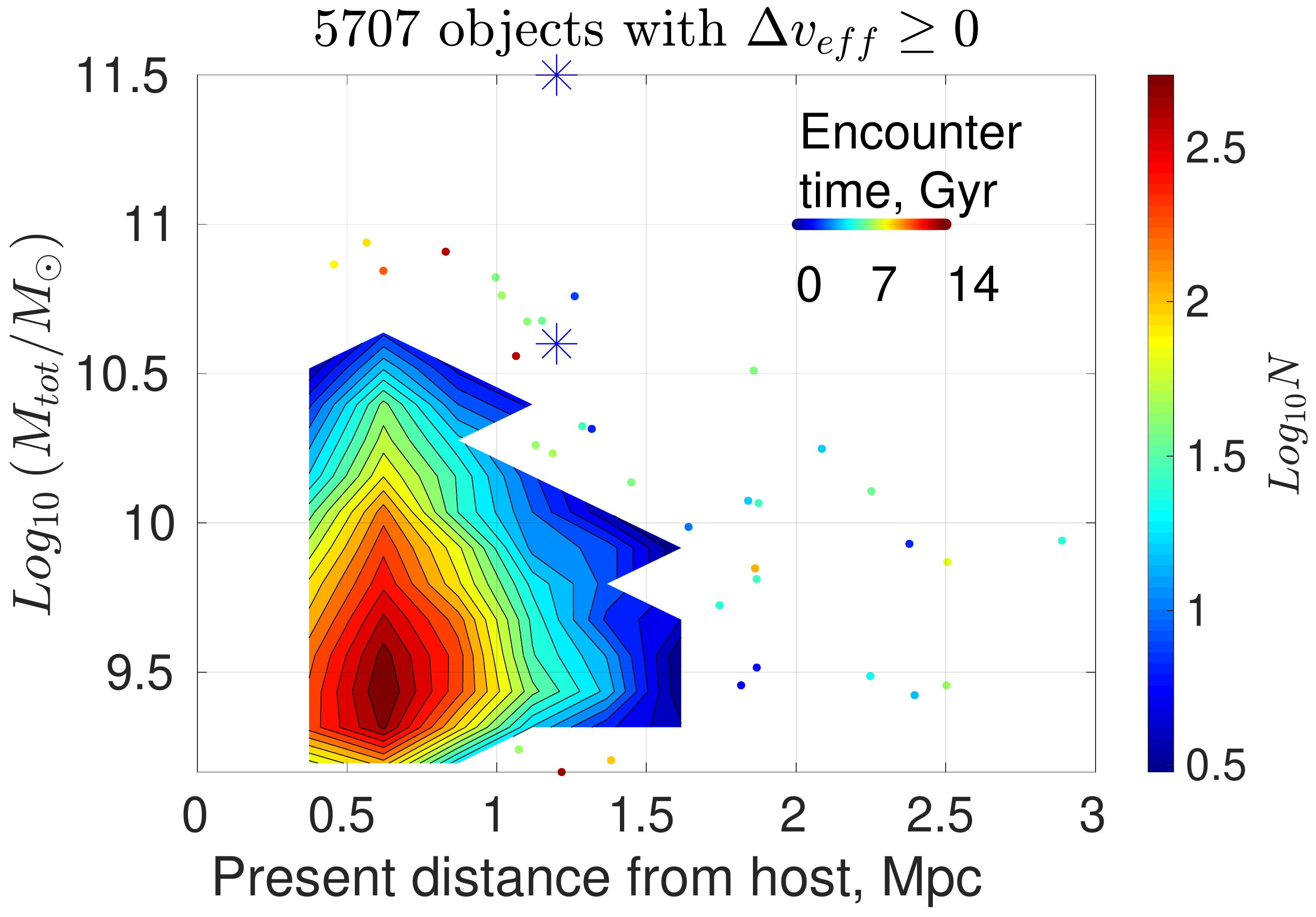}
	\caption{Similar to Figure \ref{M_vir_Delta_v_0}, but now showing results for the dark matter-only version of Illustris TNG300. Due to the much larger number of backsplashers, we again use contours to show the most densely populated regions of the mass-distance plane. Backsplashers are shown individually outside this region, with the colour of each point indicating the encounter time (similarly to Figure \ref{M_vir_Delta_v_no_limits}). The lone backsplasher with higher distance and mass than NGC~3109 (for the lower mass estimate) has $\Delta v_{eff} = 105$~km/s, which we argued in Section \ref{Requiring_energy_gain} is insufficient to explain its anomalously high RV. For clarity, we have omitted a handful of low-mass backsplashers at large distances $-$ these are much less massive than NGC~3109.}
	\label{M_vir_Delta_v_0_DMO}
\end{figure}

To check this, we compare our results with the dark matter-only version of Illustris TNG300. The analogous results to Figure \ref{M_vir_Delta_v_0} are shown in Figure \ref{M_vir_Delta_v_0_DMO}. The overall mass-distance distribution of backsplashers is quite similar in the dark matter-only run, but there are many more backsplashers in this case. This could be related to the much stronger tides upon closely approaching the host galaxy. In a hydrodynamical simulation, this would typically contain a baryonic component that is much more centrally concentrated than the dark matter. Particularly strong tides would arise if the host galaxy develops a thin disc, which can efficiently disrupt haloes passing close to it \citep{Pawlowski_2019}.

\begin{figure}
	\centering
	\includegraphics[width = 8.5cm] {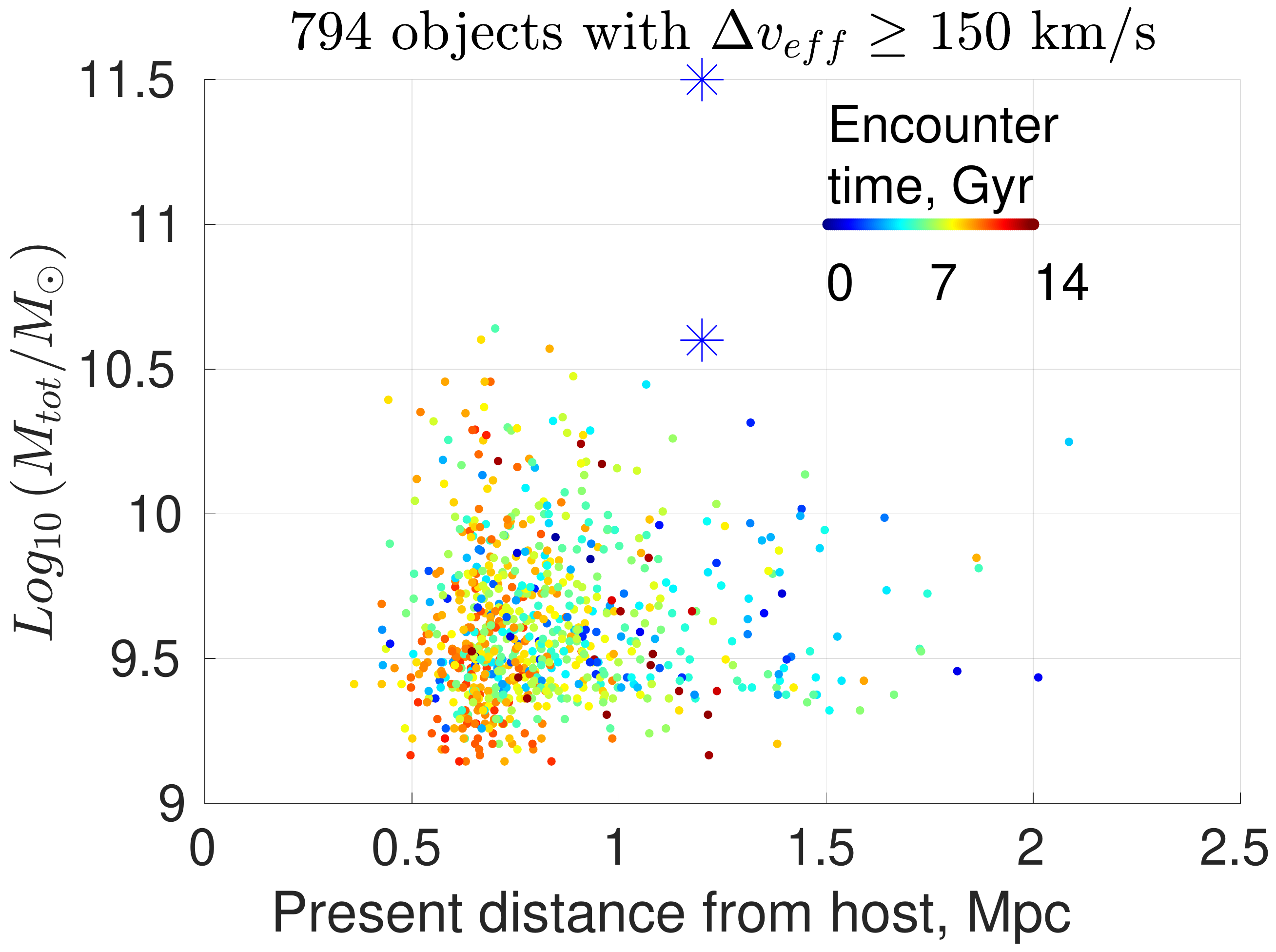}
	\caption{Similar to Figure \ref{M_vir_Delta_v_0_DMO}, but now requiring $\Delta v_{eff} \geq 150$~km/s. There are no longer enough backsplashers to reliably draw contours of their number density, so they are shown individually. Notice the rather similar result to the corresponding hydrodynamical simulation (Figure \ref{M_vir_Delta_v_150}).}
	\label{M_vir_Delta_v_150_DMO}
\end{figure}

Our results in Figure \ref{M_vir_Delta_v_0_DMO} indicate that there is one backsplasher with a marginally larger distance and mass than NGC~3109 for our very conservative choices of these parameters (Section \ref{Observed_properties_NGC3109}). However, this backsplasher has $\Delta v_{eff} = 105$~km/s, which suggests that it did not gain enough energy during the interaction to explain the anomalously high RV of NGC~3109 (Section \ref{Requiring_energy_gain}). As discussed there, a more realistic picture can be obtained by requiring $\Delta v_{eff} \geq 150$~km/s, which yields the results shown in Figure \ref{M_vir_Delta_v_150_DMO}. This demonstrates that NGC~3109 is a significant outlier also in the dark matter-only version of TNG300, with the results being rather similar to the standard hydrodynamical version used elsewhere in this contribution (Figure \ref{M_vir_Delta_v_150}). Therefore, it is clear that baryonic physics has only a small effect on our conclusion that NGC~3109 is too distant and massive to be a backsplasher from the MW or M31 in a $\Lambda$CDM context.

\section{Discussion}
\label{Discussion}

We showed that no backsplashers in the Illustris TNG300 simulation have the right mass and distance to be considered analogues of NGC~3109 even under conservative assumptions. This is consistent with the analytic estimate that backsplashers should not be found at $d \ga 2.5 \, r_{vir}$ from their host \citep{Mamon_2004}. Due to the large number of MW analogues, we are able to get some backsplashers at even larger distances. However, this is quite rare $-$ we found only 781 backsplashers with $\Delta v_{eff} \geq 0$ from 13225 host galaxies (Figure \ref{M_vir_Delta_v_0}). Some of these probably have $d < 2.5 \, r_{vir}$ as we only require $d > 2 \, r_{vir}$. Thus, our results broadly support the analytic estimate of \citet{Mamon_2004}.

%If requiring $\Delta v_{eff} > 150$~km/s, get 103 from hosts with only one associated backsplasher and the other 6 from hosts with two associated backsplashers.

For a paired host configuration, numerical simulations show that backsplashers can reach up to $\approx 5 \, r_{vir}$ from their host \citep{Ludlow_2009, Wang_2009}, broadly consistent with our results in Figure \ref{M_vir_Delta_v_0_LGlike_only} for $r_{vir} \approx 200$~kpc. This was also demonstrated in figure 1 of \citet{Teyssier_2012}. However, this figure demonstrates that the distribution of backsplashers is significantly elongated along the axis connecting the two main galaxies. In the orthogonal direction, the extent is similar to the analytic estimate of $2.5 \, r_{vir} = 660$~kpc for $M_{vir} = 2 \times 10^{12} M_\odot$. In this regard, it is worth mentioning that NGC~3109 lies 706~kpc from the MW-M31 axis (765~kpc for a distance of 1.3~Mpc), and appears on our sky in the opposite hemisphere to M31 \citep[e.g. see figure 10 of][]{Banik_Zhao_2016}. Most of the backsplashers in \citet{Teyssier_2012} are located quite close to the two main host galaxies, possibly because their combined gravity makes it difficult to reach a large distance from their barycentre. Their figure 1 shows that it is very difficult to find a backsplasher whose minimum distance from either host is 1.2~Mpc and which lies 700~kpc from the axis between the hosts.

In addition, figure 4 of \citet{Teyssier_2012} indicates that regardless of the position, backsplashers more massive than $10^{10.2} M_\odot$ are very rare. The mass of NGC~3109 is thus unusually high for a backsplasher even if we assume that its mass is $10^{10.6} M_\odot$, which is the minimum required in Newtonian rotation curve fits with plausible dark matter haloes (Figure \ref{Li_results}). Our results agree that this is unusually massive for a $\Lambda$CDM backsplasher at any distance (Figure \ref{M_vir_Delta_v_0}).

\begin{figure}
	\centering
	\includegraphics[width = 8.5cm] {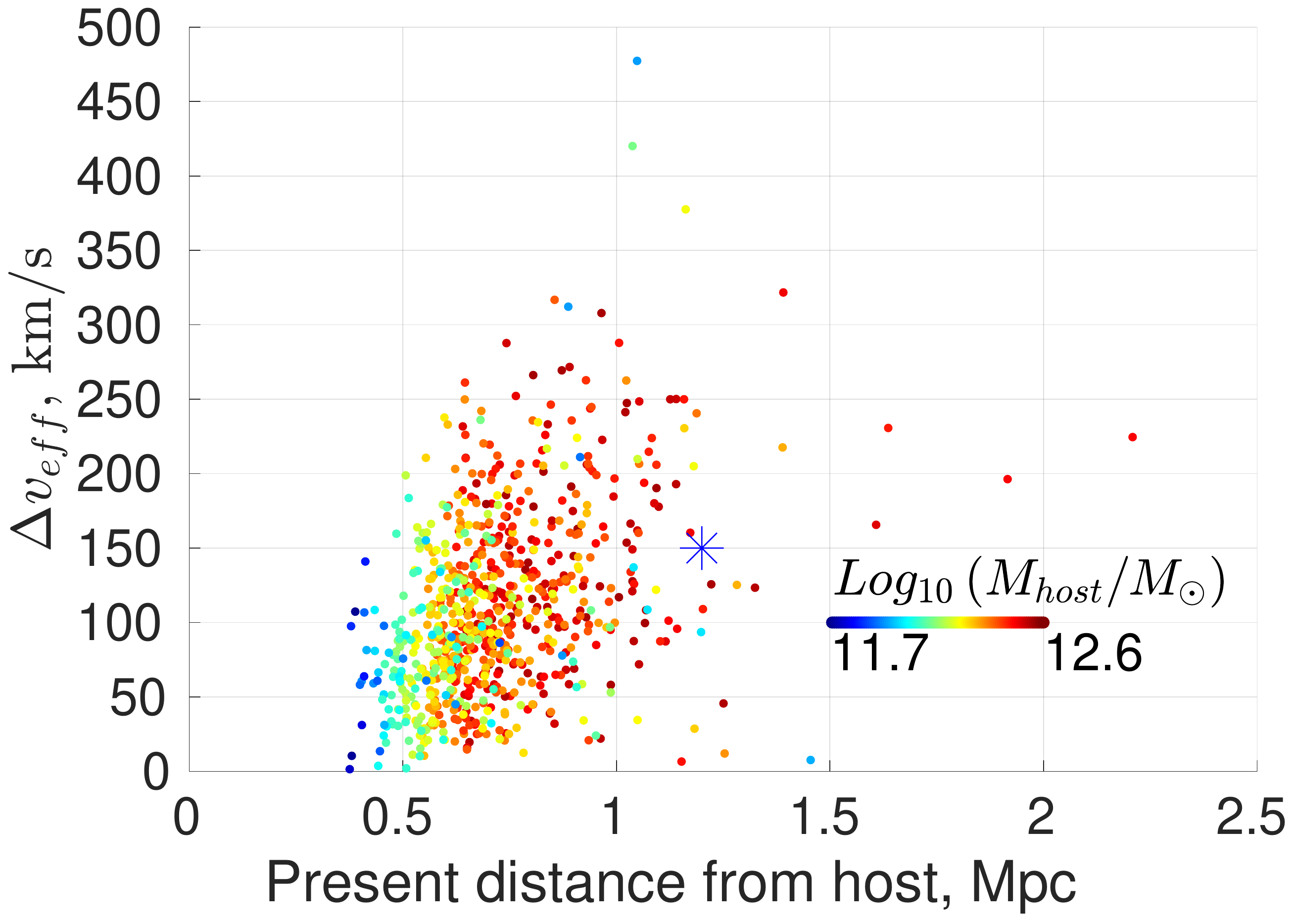}
	\caption{$\Delta v_{eff}$ as a function of present distance from the host, showing only backsplashers with $\Delta v_{eff} \geq 0$. The colours indicate the present virial mass of the host galaxy. Notice that less massive hosts preferentially appear near the bottom left, as expected for weaker backsplash events.}
	\label{r_dv}
\end{figure}

To better understand the properties of backsplashers in $\Lambda$CDM, we use Figure \ref{r_dv} to show the distribution of $\Delta v_{eff}$ and present distance from the host. As expected, backsplashers from less massive hosts are generally still quite close to their host and did not gain much energy when interacting with it. Since most of our host galaxies are more massive than $10^{12} M_\odot$, our results should not be much affected by slight adjustments to the lower limit on the allowed host mass. Changing the upper limit would have a more significant impact on the statistics, but would preferentially remove those backsplashers which get closest to reproducing the observed properties of NGC~3109.

%If we stick to such a model, we would need to assume that the galaxies in the NGC~3109 association independently fell into the virial radius of the MW/M31 and were subsequently flung out in the same direction. The very low probability of this was discussed in \citet{Banik_2018_anisotropy}. Thus, the difficulties facing $\Lambda$CDM from the mere existence of several massive distant backsplashers are further compounded when we take into account that these galaxies are rather close to each other.

\begin{figure}
	\centering
	\includegraphics[width = 8.5cm] {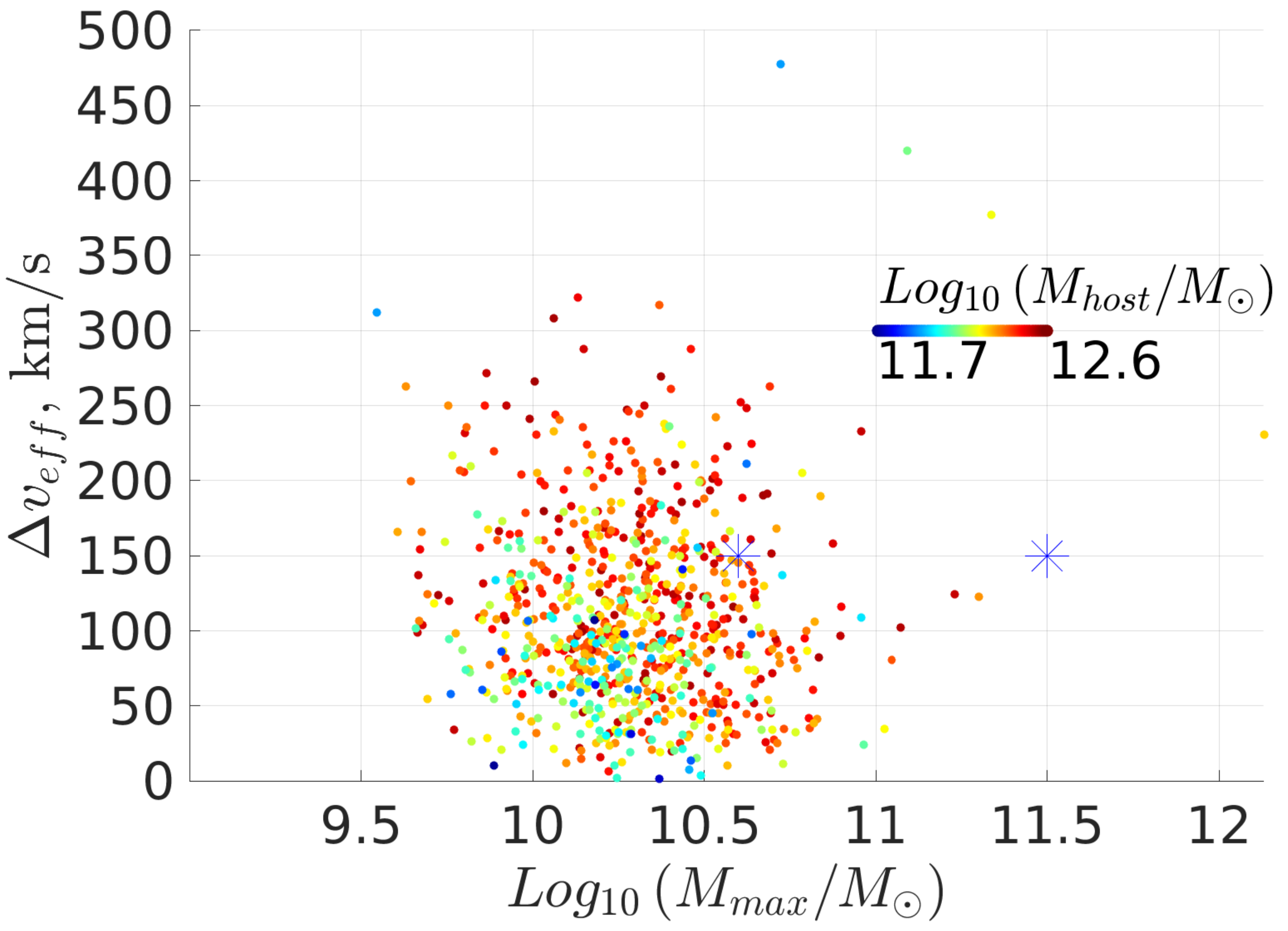}
	\includegraphics[width = 8.5cm] {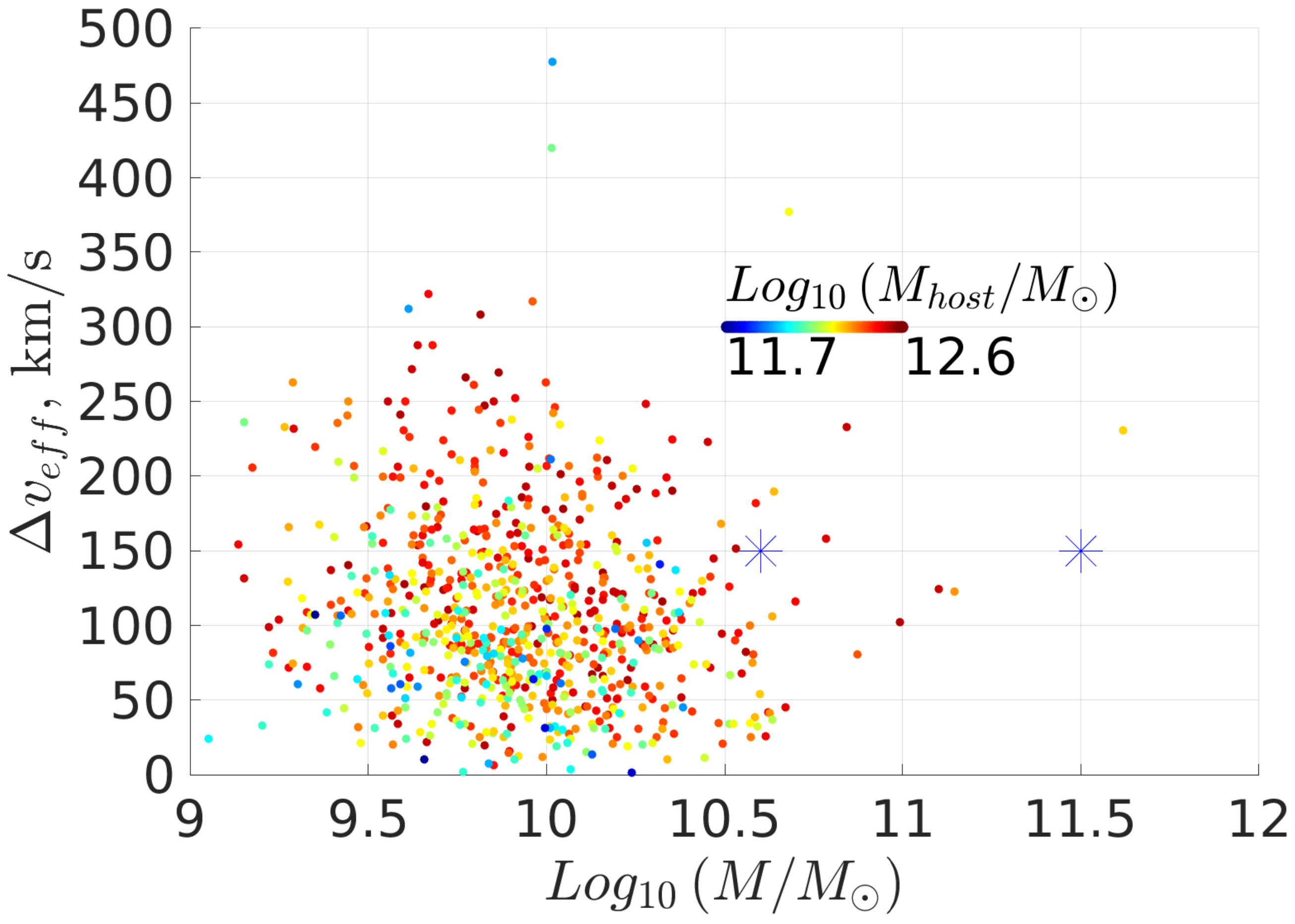}
	\caption{$\Delta v_{eff}$ as a function of maximum (\underline{top}) or present (\underline{bottom}) backsplasher total mass, showing only those objects with $\Delta v_{eff} \geq 0$. The colour indicates the present mass of the host galaxy. Notice the upper limit to $\Delta v_{eff}$ declines with mass. The most massive backsplasher in both panels is the same object, which is currently at a distance of 1.16 Mpc and thus is not analogous to NGC 3109.}
	\label{M_dv}
\end{figure}

In a $\Lambda$CDM context, an important reason for our lack of backsplash analogues to NGC~3109 is that dynamical friction would be quite significant during any past close interaction with the MW or M31 \citep{Privon_2013}. To understand the effect of dynamical friction, \citet{Teyssier_2012} used their figure 4 to show how the mass distribution of backsplashers changes with distance from the host. The statistics were limited as only one LG-like host was considered. We revisit this issue in a slightly different way using our sample of 13225 host galaxies. The energy gain $\Delta v_{eff}$ should be smaller at higher mass due to the effect of dynamical friction. Thus, we use Figure \ref{M_dv} to show the relation between $\Delta v_{eff}$ and backsplasher mass for the cases where $\Delta v_{eff} \geq 0$, with the host mass indicated using the colour. As expected, the upper limit to $\Delta v_{eff}$ declines with mass. There are very few backsplashers more massive than NGC~3109 with $\Delta v_{eff} \geq 150$~km/s, which we argued is required to reach 1.2~Mpc from the host (Section \ref{Requiring_energy_gain}). The handful of such massive backsplashers all have $d < 1.2$~Mpc (Figure \ref{M_vir_Delta_v_150}), even if we relax the energy gain requirement to a very conservative $\Delta v_{eff} \geq 0$ (Figure \ref{M_vir_Delta_v_0}).

Another reason for this lack of backsplash analogues to NGC~3109 could be that backsplashers lose mass during their encounter with a more massive galaxy \citep[e.g.][]{Smith_2016}. This is apparent by comparing the top and bottom panels of our Figure \ref{M_dv}, which show the maximum and present mass, respectively, of each backsplasher. The loss of mass during the encounter is also apparent upon closer examination of the only trajectory shown in Figure \ref{Illustris_trajectories} where the backsplasher significantly gained energy during the interaction. The high $\Delta v_{eff}$ of $225$~km/s comes at the price of the backsplasher mass decreasing from $3.14 \times 10^{10} M_\odot$ at $t_{in}$ to $1.48 \times 10^{10} M_\odot$ at $t_{out}$. Neither mass would be enough to explain the observed rotation curve of NGC~3109, but the situation is substantially worse post-encounter.

As discussed in Section \ref{NGC3109_mass}, our adopted mass for NGC~3109 should also consider the rest of the NGC~3109 association. While it may well be unbound today \citep{Kourkchi_2017}, it should have been bound in the past in order to explain the filamentary nature of the NGC~3109 association \citep{Bellazzini_2013}. Moreover, the other galaxies in this association are likely also backsplashers since their RVs are also too high to be accounted for by the timing argument analyses of \citet{Peebles_2017} and \citet{Banik_2018_anisotropy}. Although the Illustris simulation may well struggle to resolve galaxies like Leo P, our results strongly suggest that it would be very difficult to find a backsplasher of any mass at its observed distance of $1.62 \pm 0.15$~Mpc \citep{McQuinn_2015}. Even if this were possible, we would have to explain why several unassociated dwarf galaxies should be flung away from the LG in much the same direction and end up at a similar distance, suggesting a similar encounter time with the host galaxy. If these dwarfs were falling along a filament onto the MW while it was undergoing a minor merger, then the short dynamical timescale of the merger implies that even slight differences in the infall time of the dwarf could substantially alter the direction in which it is ultimately flung out. Moreover, galaxies with different mass should have experienced different amounts of dynamical friction. This means that even if several galaxies fell into the MW along a filament at much the same time and were ejected outwards in a similar direction, they would end up at rather different distances, e.g. Sextans A would still end up much further ahead than NGC~3109. In reality, both have a similar distance, with Sextans A only $\approx 12\%$ further away \citep{McQuinn_2017}.%Thus, $\Lambda$CDM does not admit a correlated origin for the NGC~3109 galaxies if one considers them to be backsplash. Previous works showed that assuming otherwise contradicts the timing argument, since in this case they would never have been very close to a major LG galaxy and thus a 3D few-body model of the LG should be sufficient \citep{Banik_2018_anisotropy, Peebles_2017}.

It is also worth mentioning that essentially all the HVGs identified by \citet{Banik_2018_anisotropy} lie in the NGC~3109 association, even though they considered 33 galaxies in addition to the MW and M31. Their figure 10 shows that there are at best three other HVGs in addition to those in the NGC~3109 association. Of these, the distance to HIZSS 3 is seriously questionable due to observational difficulties caused by its extremely low Galactic latitude of $0.09^\circ$ \citep[section 6.3 of][]{Banik_2018_anisotropy}. Meanwhile, the RVs of KKH 98 and DDO 190 are marginally consistent with the dynamical model if we allow a model uncertainty of 25~km/s, which is also suggested by focusing on only those galaxies whose RVs lie below the model prediction. Thus, postulating that the NGC~3109 association was never bound would still leave us with the task of explaining the anomalously high RVs of NGC~3109, Sextans A, Sextans B, and very probably Leo P. Whatever process is responsible for these HVGs, it is not very common. In the relatively small fraction of cases where the unknown process operates, the resulting HVGs should not end up in the same direction at a similar distance if the HVGs were flung out in individual events. This was discussed in great detail by \citet{Banik_2018_anisotropy}, who suggested that the HVGs must have been correlated in some way based purely on how they define a thin plane. A correlation becomes almost inevitable when we consider that most if not all of the HVGs are actually located quite close to a line \citep{Pawlowski_McGaugh_2014}.

The most plausible solution is that the NGC~3109 association was once a gravitationally bound group. The mass required to bind it would be rather large, with \citet{Bellazzini_2013} estimating that the required mass was $10^{11.5} M_\odot$. Such a high mass could well alleviate the above-mentioned issues regarding the NGC~3109 association, but would also increase the amount of dynamical friction during any close encounter with a massive galaxy. Our results in Figure \ref{M_vir_Delta_v_150} show that a backsplasher of this mass is highly implausible in a model where galaxies have dark matter haloes that would inevitably create significant dynamical friction during interactions \citep{Privon_2013, Kroupa_2015}. Since there are no analogues to NGC~3109 for an assumed mass of just $10^{10.5} M_\odot$, it is clear that the galaxy and the rest of its association pose severe problems for $\Lambda$CDM if their high RVs indicate that they are backsplash from the MW or M31, as argued here and in previous works \citep{Teyssier_2012, Pawlowski_McGaugh_2014, Banik_2018_anisotropy}.

So far, we have mostly focused on comparing backsplashers to the present mass of NGC~3109. However, our preceding discussion suggests that it should have been much more massive in the past to bind the NGC~3109 association. Since mass could be lost during a past encounter with the MW, a conservative approach would be to consider the maximum mass of each backsplasher at any time in its past, as traced by the Illustris merger tree. This is shown in the bottom panel of Figure \ref{M_dv} against the backsplasher's $\Delta v_{eff}$. It is evident that only one backsplasher with $\Delta v_{eff} \geq 150$~km/s ever had a mass $\geq 10^{11.5} M_\odot$, but it is too close to its host to resemble NGC 3109.

\subsection{Broader context: the satellite planes challenge}
\label{Broader_context}

The HVGs in the NGC~3109 association should be considered together with the LG satellite planes because these could all have a common origin, as suggested in Section \ref{MOND_scenario}. Indeed, \citet{Pawlowski_McGaugh_2014} showed that the high RVs of galaxies in the NGC 3109 association strongly suggest a past close interaction with the MW, even though the association currently lies outside the zero-velocity surface of the LG. For this reason, we do not combine the probabilities of these challenges, but do consider the level of tension with $\Lambda$CDM.

Flattened distributions of very likely co-orbiting satellite galaxies around the MW \citep{Lynden_Bell_1976, Lynden_Bell_1982, Kroupa_2005, Pawlowski_2012} and M31 \citep{Metz_2007, Ibata_2013} have long posed a challenge to our understanding of galaxy formation in the $\Lambda$CDM context. Recent proper motion data confirm that most of the classical MW satellites do indeed have a common orbital plane \citep{Pawlowski_2013, Pawlowski_2020} aligned with the plane normal defined by the satellite positions alone \citep{Isabel_2020}. Their velocities show a very significant bias towards the tangential direction, as occurs for a rotating disc \citep{Cautun_2017}. Proper motions of two M31 satellite plane members indicate that this structure is likely also coherently rotating \citep{Sohn_2020}, as suggested by the RVs of satellites in this nearly edge-on structure \citep{Ibata_2013}. After careful consideration of several proposed scenarios for how primordial CDM-rich satellites might end up in a thin plane, \citet{Pawlowski_2014} concluded that none of them agree with observations for either the MW or M31. Structures as extreme as those observed are exceedingly rare in cosmological simulations \citep{Ibata_2014, Pawlowski_2014_paired_halos}, including hydrodynamical simulations \citep{Ahmed_2017, Shao_2019, Pawlowski_2020} and simulations which model the effects of a central disc galaxy \citep{Pawlowski_2019}. The arguments raised by \citet{Metz_2009_infall} and \citet{Pawlowski_2014} against the group infall and filamentary accretion scenarios were later independently confirmed by \citet{Shao_2018} using the EAGLE hydrodynamical cosmological simulation \citep{Crain_2015, Schaye_2015}. For a recent review of the satellite plane problem, we refer the reader to \citet{Pawlowski_2018}, who considered both LG satellite planes and the recently discovered one around Centaurus A \citep{Muller_2018, Muller_2021}.

To help our discussion, we quantify the level of tension that each challenge represents for $\Lambda$CDM, and compare to the one found here. Since we found no NGC~3109 analogues around 13225 hosts, we conservatively assign a frequency of 1/13225 to the HVG challenge. The equivalent number of standard deviations $\chi$ corresponding to this frequency can be found using
\begin{eqnarray}
    1 - \frac{1}{\sqrt{2 \mathrm{\pi}}} \int_{-\chi}^\chi \exp \left( -\frac{x^2}{2} \right) \, dx ~=~ P \, .
    \label{P_chi}
\end{eqnarray}
We solve this using the Newton-Raphson algorithm with initial guess $\left( 3 - \log_{10} P \right)$ for events with $P < 0.001$ (the tension is not very significant otherwise, making the initial guess less important for numerical convergence). In this way, we estimate that the HVG challenge corresponds to a $3.96\sigma$ event.

This is based on allowing the LG mass to lie in the range $\left( 2 - 5 \right) \times 10^{12} M_\odot$, with consequent implications for the allowed mass range of isolated hosts (Section \ref{Isolation_conditions}) and the isolation condition on backsplashers (Section \ref{Backsplasher_isolation}). If the LG mass is restricted to the range $\left( 2 - 4 \right) \times 10^{12} M_\odot$ by reducing $M_{max}$ in Equation~\ref{Max_mass_ratio}, the slightly reduced number of hosts raises the frequency to 1/12187 ($3.94\sigma$). If instead we restrict the LG mass to $\left( 3 - 5 \right) \times 10^{12} M_\odot$ by raising $M_{min}$ while keeping the nominal $M_{max}$, the number of hosts decreases to 10089, reducing the significance to $3.89\sigma$. Using the nominal LG mass range of $\left( 2 - 5 \right) \times 10^{12} M_\odot$ but focusing on our dark matter-only simulation, we get 12960 hosts, yielding a significance of $3.95\sigma$. In all these cases, the estimated statistical significances should be treated as lower limits because we did not identify any backsplash analogues to NGC~3109.

\begin{table}
	\centering
	\begin{tabular}{ccc}
		\hline
		Problem for $\Lambda$CDM & Frequency & Significance \\ \hline
		MW satellite plane & 1/2548 & $3.55\sigma$ \\
		M31 satellite plane & 3/7757 & $3.55\sigma$ \\
		NGC 3109 backsplash (this work) & 1/13225 & $3.96\sigma$ \\ \hline
	\end{tabular}
	\caption{The level of tension between $\Lambda$CDM and various challenges it faces within the LG. We have chosen these challenges because they have a common explanation in an alternative framework (Section \ref{MOND_scenario}). The frequencies for the MW and M31 satellite planes come from section 4.2 of \citet{Pawlowski_2020} and figure 2 of \citet{Ibata_2014}, respectively. Values in the final column are obtained by applying Equation \ref{P_chi} to the frequencies.}
	\label{Statistical_significance}
\end{table}

Table \ref{Statistical_significance} summarizes the statistical significance of the HVG challenge for $\Lambda$CDM and that of the LG satellite planes. These probabilities are a frequentist interpretation of the ``number of trials'' for each individual test \citep[e.g.][]{Bayer_2020}. Since the timing argument analysis of \citet{Banik_2018_anisotropy} considered the kinematics of 33 LG non-satellite dwarf galaxies of which NGC 3109 was not the only HVG, the challenge to $\Lambda$CDM posed by NGC 3109 is difficult to understand merely via the look-elsewhere effect. Moreover, only around the MW, M31, and Centaurus A do we have information on the 3D distribution of satellite galaxies and at least one component of their velocity. A satellite plane is also evident around Centaurus A, with properties that are likely to arise in $\Lambda$CDM only 0.2\% of the time \citep{Muller_2021}. As a result, it would be difficult to repeat the above-mentioned analyses further afield. In particular, distance uncertainties would be larger for more distant objects, creating significant uncertainty on the peculiar velocity and making it very tricky to do a timing argument analysis. Distance uncertainties also make it difficult to determine the 3D distribution of satellite galaxies, which in addition are very faint and not easy to observe even at the distance of Centaurus A. Thus, the above-mentioned challenges for $\Lambda$CDM arise in the only cases where the paradigm can be tested in detail based on the timing argument and the phase space distribution of satellites. One can consider other tests of $\Lambda$CDM beyond the LG, some of which we briefly discuss next.

\subsection{An alternative scenario}
\label{MOND_scenario}

Anisotropically distributed satellite galaxies are known to form out of the tidal debris expelled during the interaction between galaxies, as observed in the Antennae \citep{Mirabel_1992}. Therefore, the MW and M31 co-orbiting planes of satellite galaxies may have formed as tidal dwarf galaxies \citep[TDGs;][]{Pawlowski_2011}, implying a past major interaction. The HVGs in the NGC~3109 association may then be backsplash from this event, even if backsplash events with the required properties do not occur in $\Lambda$CDM.

Any second-generation origin for the MW and M31 satellite planes runs into the issue that such satellites would be free of dark matter \citep{Barnes_1992, Wetzstein_2007, Haslbauer_2019} $-$ this was discussed in more detail by \citet{Kroupa_2012}. But since CDM haloes have never been detected independently of their presumed gravitational effects \citep[e.g.][]{Hoof_2020} and require particles beyond the well-tested standard model of particle physics, it is prudent to consider alternative paradigms without such haloes \citep[e.g.][]{Kroupa_2015}. The most promising such paradigm is Milgromian dynamics \citep[MOND;][]{Milgrom_1983}. In MOND, the gravitational field strength $g$ at distance $r$ from an isolated point mass $M$ transitions from the Newtonian ${GM/r^2}$ law at short range to
\begin{eqnarray}
	g ~=~ \frac{\sqrt{GMa_{_0}}}{r} ~~~\text{for } ~ g \ll a_{_0} \, .
	\label{Deep_MOND_limit}
\end{eqnarray}
MOND introduces $a_{_0}$ as a fundamental acceleration scale of nature below which the deviation from Newtonian dynamics becomes significant. Empirically, $a_{_0} \approx 1.2 \times {10}^{-10}$ m/s$^2$ to match galaxy rotation curves \citep{Begeman_1991, Gentile_2011}. With this value of $a_{_0}$, MOND continues to fit galaxy rotation curves very well using only their directly observed baryonic matter \citep[e.g.][]{Kroupa_2018, Li_2018, Sanders_2019}. In particular, observations confirm the prior MOND prediction of very large departures from Newtonian dynamics in low surface brightness galaxies \citep[e.g.][]{Blok_1997, McGaugh_1998}. More generally, there is a very tight empirical `radial acceleration relation' (RAR) between the gravity inferred from rotation curves and that expected from the baryons alone in Newtonian dynamics \citep{McGaugh_Lelli_2016}, with the relation also extending to ellipticals \citep{Lelli_2017, Chae_2020_elliptical, Shelest_2020}. This confirms the central prediction of \citet{Milgrom_1983}.

The evidence for MOND on galaxy scales goes beyond the observed tightness of the RAR. For instance, the dynamical friction experienced by galactic bars rotating through a CDM halo is problematic because it would cause the bar to slow down \citep{Debattista_2000}, conflicting with observations \citep{Algorry_2017, Peschken_2019} $-$ the tension is at the $8\sigma$ level \citep{Roshan_2021}. In addition, bar-halo angular momentum exchange would cause a resonant effect leading to a quite strong bar after only a few Gyr \citep{Athanassoula_2002}, making it difficult to explain rather isolated galaxies like M33 with only a weak bar \citep{Sellwood_2019}. This is naturally accounted for in a hydrodynamical MOND simulation of M33, which bears good overall resemblance to observations \citep{Banik_2020_M33}. The lack of massive CDM haloes and the resulting dynamical friction in close interactions causes a reduced major merger rate \citep{Nipoti_2007, Renaud_2016}, which might better explain the high prevalence of thin disc galaxies in the local Universe with little or no bulge \citep{Kormendy_2010, Peebles_2010}. This continues to challenge the latest $\Lambda$CDM cosmological simulations \citep{Peebles_2020}. For a review of MOND including its strengths and weaknesses, we refer the reader to \citet{Famaey_McGaugh_2012}, while \citet{Milgrom_2015} provides a more theoretical review.

In the LG, Equation \ref{Deep_MOND_limit} implies a much stronger MW-M31 mutual attraction than the Newtonian inverse square law. As a result, applying MOND to the almost radial MW-M31 orbit \citep{Van_der_Marel_2012, Van_der_Marel_2019, Salomon_2021} implies that they underwent a close encounter ${9 \pm 2}$~Gyr ago, as first put forward by \citet{Zhao_2013}. This is approximately when the MW bar formed and its disc underwent the buckling instability \citep{Grady_2020}, which could be due to the interaction if it was $8-9$~Gyr ago. Due to the high MW-M31 relative velocity around the time of their flyby, they would likely have gravitationally slingshot several LG dwarfs out at high speed. This could well explain the unusually high RV of NGC~3109 $-$ it might have been near the spacetime location of the flyby, thereby gaining a significant amount of energy from the time-dependent LG potential \citep{Banik_2018_anisotropy}. Their figure 5 shows that backsplashers from such a highly energetic flyby can easily reach the 1.3~Mpc distance of NGC~3109, and even the 1.6~Mpc distance of Leo P. This is because in addition to the fast MW-M31 relative velocity of $\approx 700$~km/s at pericentre, dynamical friction would be greatly reduced as galaxies would not have dark matter haloes \citep{Bilek_2018}. %Nonetheless, the enhanced long range gravity in MOND allows the MW-M31 timing argument to be satisfied with just their baryonic mass \citep[section 2 of][]{Banik_Ryan_2018}.

In this scenario, NGC~3109 must have closely approached the MW and/or M31. However, figure 6 of \citet{Banik_2018_anisotropy} shows that it is quite possible for the MW-M31 interaction to efficiently slingshot a tracer particle out to ${> 1.6}$~Mpc even if it never approached within 40 disc scale lengths of either galaxy. Thus, it is easy to envisage the NGC~3109 association being flung out in this way with negligible dynamical friction on its constituents. It is likely that the association as a whole would be tidally disrupted, such that it is likely unbound today \citep{Kourkchi_2017}. This would make the association analogous to a tidal stream traced by dwarf galaxies rather than stars. However, tidal effects on individual galaxies in the association may have been rather small due to the large pericentric distance and the short duration of any such interaction. Even if there were tidal signatures imprinted at pericentre, the long time since then would make it nearly impossible to identify them today.

During the MW-M31 flyby, tidal tails would likely have formed and might later have condensed into TDGs \citep{Zhao_2013}. This phenomenon occurs in some observed galactic interactions like the Antennae \citep{Mirabel_1992} and in MOND simulations of them \citep{Tiret_2008, Renaud_2016}. Due to the way in which such TDGs form out of a thin tidal tail, they would end up lying close to a plane and co-rotating within that plane \citep{Wetzstein_2007, Haslbauer_2019}, though a small fraction might well end up counter-rotating depending on the exact details \citep{Pawlowski_2011}. The possibility of explaining the LG satellite planes in this way was investigated with MOND $N$-body simulations of the MW-M31 encounter \citep{Bilek_2018}. Those authors demonstrated the formation of tidal tails connecting the galaxies. \citet{Banik_Ryan_2018} investigated a much wider range of orbital geometries using a restricted $N$-body approach where the MW and M31 were treated as point masses surrounded by test particle discs. The tidal debris around each galaxy were generally distributed in a thin plane, as evidenced by a sharp concentration of orbital poles. In some models, the preferred direction aligned with the corresponding observed satellite plane for both the MW and M31 (see their figure 5). One reason for this success is that the MW and M31 satellite planes rotate in the same sense, with their orbital poles separated by only $\approx 50^\circ$ \citep{Pawlowski_2014}. While some mismatch is expected due to the orientations of the MW and M31 discs differing by $\approx 65^\circ$ \citep[table 1 of][]{Banik_2018_anisotropy}, a much larger angle would be difficult to accommodate if both satellite planes condensed out of a common tidal tail. %We are currently conducting hydrodynamical simulations of this interaction using Phantom of \textsc{RAMSES} \citep[\textsc{por},][]{Lughausen_2015}, an adaptation of the \textsc{ramses} algorithm widely used in astronomy \citep{Teyssier_2002}. 

Since the encounter would have been very long ago, the metallicities and other internal properties of the M31 satellite plane members might be rather similar to those of primordial dwarfs \citep{Recchi_2015}, especially in a model where both TDGs and primordial dwarfs are purely baryonic and thus lack any fundamental difference. This might explain the similarity in internal properties between on- and off-plane satellites of M31 \citep{Collins_2015}. While those authors interpreted their results as evidence against the TDG hypothesis, field dwarfs (which are presumably mostly primordial in a $\Lambda$CDM context) follow a similar mass-radius relation to confirmed TDGs \citep{Dabringhausen_2013}. However, a clear splitting is expected in cosmological $\Lambda$CDM simulations \citep[figure 12 of][]{Haslbauer_2019}. The similarity between primordial and tidal dwarfs is expected in MOND as both would be purely baryonic.

Therefore, the MOND scenario of a past MW-M31 flyby could well explain the LG satellite planes and the high internal velocity dispersions of their members while also accounting for the unusually high RV of NGC~3109 for its position. This should be seriously considered as an alternative to the standard approach of treating the HVG and satellite plane problems as separate statistical flukes in the $\Lambda$CDM paradigm (Table \ref{Statistical_significance}). It should be the topic of further detailed simulations in a MOND context.

\section{Conclusions}
\label{Conclusions}

A detailed 3D Newtonian timing argument calculation of the LG and its surroundings underpredicts the RV of NGC~3109 by $105 \pm 5$~km/s \citep[table 3 of][]{Banik_2018_anisotropy}. This is despite the significantly more exhaustive search through parameter space described in their section 4.1 compared to the similar analysis of \citet{Peebles_2017}, who reached similar conclusions. No simple trajectory can be found for these galaxies that respects the Newtonian timing argument and matches available observations at $z = 0$.

However, the analyses of \citet{Peebles_2017} and \citet{Banik_2018_anisotropy} are not cosmological simulations. In such a simulation, there could be processes which are not correctly handled by the above-mentioned analyses. In particular, mergers between galaxies can temporarily lead to high relative velocities. A nearby dwarf could then be flung outwards at high speed, possibly explaining the anomalously high RV of NGC~3109. This would make it a backsplasher, as previously suggested by \citet{Teyssier_2012} and \citet{Pawlowski_McGaugh_2014}. Using a simplified calculation, we found that this scenario requires an energy gain of $\Delta v_{eff} \ga 150$~km/s during a past interaction with the MW (Section \ref{Requiring_energy_gain}). Such trajectories can increase the RV by $\approx 110$~km/s compared to a $\Delta v_{eff} = 0$ trajectory that reaches the same present Galactocentric distance of 1.2~Mpc. Thus, backsplash can in principle explain the anomalously high RV of NGC~3109.

To find out if such trajectories are expected in $\Lambda$CDM, we investigated the Illustris TNG300 hydrodynamical cosmological simulation. We identified 13225 host galaxies similar to the MW or M31, and used the merger tree to trace them back in time. At each snapshot, we identified all subhaloes within their virial volume, and traced them forwards as far as possible. Backsplashers are those with a recognizable root descendant at the present epoch that lies beyond $2 \, r_{vir}$ from the associated host (Section \ref{Backsplash_condition}).

We found that backsplashers with a larger distance and mass than NGC~3109 are very rare. In the handful of cases where they do occur, $\Delta v_{eff} < 0$, probably due to dynamical friction. These backsplashers must have received a significant helping hand from large scale structure to reach their present distance, since during the encounter they actually \emph{lost} energy. However, the timing argument analyses of \citet{Peebles_2017} and \citet{Banik_2018_anisotropy} include the major galaxy groups outside the LG up to a distance of almost 8~Mpc \citep[table 3 of][]{Banik_Zhao_2017}. Therefore, the Illustris cosmological simulation does not reveal trajectories with NGC~3109-like final states that might be significantly mis-modelled by the above-mentioned 3D timing argument analyses. As these neglect dynamical friction, including this process would if anything make it even more difficult to explain the high RV of NGC~3109.

Since the backsplash process concerns the motions of fairly massive galaxies with significant CDM haloes, it should not be affected much by baryonic physics in galaxies. We tested this by comparing our results to the dark matter-only version of TNG300 (Section \ref{Dark_matter_only}). There were many more backsplashers in this case, perhaps due to the lack of strong disruptive tides from e.g. a baryonic disc in the host \citep{Pawlowski_2019}. Nonetheless, we found no backsplash analogues to NGC~3109 in the dark matter-only simulation, which yielded a similar overall distance-mass distribution for backsplashers compared to the hydrodynamical TNG300. We therefore conclude that this distribution does not extend to the observed properties of NGC~3109 regardless of precisely how the baryonic physics is treated.

To explain the anomalous kinematics of the NGC~3109 association via the backsplash process, we would need several backsplashers to be flung out in nearly the same direction at a similar time. This strongly suggests that the whole association was once a bound group which closely approached the MW or M31 and was subsequently flung out \citep{Bellazzini_2013}. The high mass required for the NGC~3109 group in this scenario renders it infeasible in the $\Lambda$CDM context because of the inevitable very strong dynamical friction during the encounter. This is apparent in the lack of sufficiently massive and distant backsplashers in the Illustris TNG300 simulation (Figure \ref{M_vir_Delta_v_0}). The situation remains the same if we trace each backsplasher back in time and consider its maximum mass (bottom panel of Figure \ref{M_dv}).

%NGC 1400 and NGC 1407 \citep{Su_2014}. But it seems to work all right in $\Lambda$CDM.

Our null detection of backsplash analogues to NGC~3109 allows us to place an upper limit on their frequency of 1/13225, implying $\Lambda$CDM is in $>3.96 \sigma$ tension with the observed properties of NGC~3109 if it is a backsplasher (Section \ref{Broader_context}). We argue that this is more probable than a 105~km/s error in the timing argument analysis of \citet{Banik_2018_anisotropy} for an isolated dwarf galaxy 1.3~Mpc away (Section \ref{Timing_argument_reliability}) that is also quite far from any major galaxy outside the LG (Table~\ref{Perturber_distances}). This problem may be related to the phase space correlated distribution of satellite galaxies around the MW and M31, each of whose satellite planes are in $3.55 \sigma$ tension with $\Lambda$CDM (Table \ref{Statistical_significance}). These should also be combined with the severe tensions that $\Lambda$CDM faces on cosmological scales with regards to the locally measured expansion rate \citep{Riess_2020, Valentino_2021}, the unusually low matter density within 300~Mpc \citep{Keenan_2013, Haslbauer_2020}, and the too-rapid formation of observed galaxy clusters like El Gordo \citep{Asencio_2021}.

We therefore propose an alternative scenario in which the unusual kinematics of the NGC~3109 association might bear witness to a past close MW-M31 flyby in the MOND context (Section \ref{MOND_scenario}). Tidal debris from the flyby could have formed into the LG satellite planes \citep{Banik_Ryan_2018, Bilek_2018}. Fitting this picture into a broader cosmological context \citep[as suggested by][]{Haslbauer_2020} would require a relativistic MOND theory such as that of \citet{Skordis_2019}, which may well enhance the growth of structure sufficiently to address the above-mentioned issues.

\section*{Data availability}

The data underlying this article are available in the article. The algorithms used to download and analyze the Illustris simulations have been made publicly available.\footnote{\href{https://seafile.unistra.fr/d/6b09464443da478d8926/}{https://seafile.unistra.fr/d/6b09464443da478d8926/}}

\section*{Acknowledgements}

IB is supported by an Alexander von Humboldt Foundation postdoctoral research fellowship. MSP and BF thank the Deutscher Akademischer Austauschdienst for PPP grant 57512596 funded by the Bundesministerium f\"ur Bildung und Forschung, and the Partenariat Hubert Curien (PHC) for PROCOPE project 44677UE. MSP thanks the Klaus Tschira Stiftung and German Scholars Organization e.V. for support via a Klaus Tschira Boost Fund. BF acknowledges funding from the Agence Nationale de la Recherche (projects ANR-18-CE31-0006 and ANR-19-CE31-0017) and from the European Research Council (ERC) under the European Union's Horizon 2020 Framework Programme (grant agreement number 834148). The authors are grateful to the referee for detailed constructive comments on the article. The graphs were produced using \textsc{matlab}$^\text{\textregistered}$.
%Thank referee.

\bibliographystyle{mnras}
\bibliography{BSP_bbl}
\bsp
\label{lastpage}
\end{document}